\documentclass[12pt,preprint]{aastex}

\shorttitle{Ensembles of Transient Cores}
\shortauthors{Garrod et al.}

\begin{document}

\title{Molecular Clouds as Ensembles of Transient Cores}

\author{R. T. Garrod,\altaffilmark{1,2} D. A. Williams,\altaffilmark{2} and 
J. M. C. Rawlings\altaffilmark{2}}
\email{rgarrod@mps.ohio-state.edu}

\altaffiltext{1}{Department of Physics, The Ohio State University, 
191 West Woodruff Avenue, Columbus, OH 43210.}
\altaffiltext{2}{Department of Physics and Astronomy, University College 
London, Gower Street, London, WC1E 6BT, UK.}

\begin{abstract}
We construct models of molecular clouds that are considered as ensembles
of transient cores. Each core is assumed to develop in the background gas
of the cloud, grow to high density and decay into the background. The
chemistry in each core responds to the dynamical state of the gas and to
the gas-dust interaction. Ices are deposited on the dust grains in the
core's dense phase, and this material is returned to the gas as the core 
expands to low density. The cores of the ensemble number typically one 
thousand and are placed randomly in position within the cloud, and are 
assigned a random evolutionary phase.

The models are used to generate molecular line contour maps of a typical
dark cloud. These maps are found to represent extremely well the
characteristic features of observed maps of the dark cloud L673, which has
been observed at both low and high resolutions. The computed maps are
found to exhibit the general morphology of the observed maps, and to
generate similar sizes of emitting regions, molecular column densities,
and the separations between peaks of emissions of various molecular
species. The models give insight into the nature of molecular clouds and
the dynamical processes occurring within them, and significantly constrain 
dynamical and chemical processes in the interstellar medium.
\end{abstract}


\section{Introduction}
There is now a large body of evidence showing that dark
molecular clouds in the interstellar medium are clumpy on a scale not
resolved by single-dish molecular line observations. For example, 
\citet{peng98a} showed from CO line observations that the
well-studied TMC-1 Core D, previously regarded as a single entity in
dynamics and chemistry, consists of some 47 distinct sub-cores. This
notable discovery has beeen shown to have implications for the evolution
of TMC-1 and its chemical richness \citep{hartquist01a}.
\citet{morata03a} used the BIMA interferometer to make a detailed study 
of structure within the molecular cloud L673. Their maps in lines of CS,
N$_2$H$^+$, and HCO$^+$ showed considerable structure not previously revealed
in single-dish studies \citep{morata97a}. This structure is different in
each molecular line, suggesting that chemical time-dependence is
responsible, and that the density inhomogeneities (hereafter described as
cores) are probably growing from low to high density on a time-scale
around one million years and then dissipating on a similar time-scale.
Further work by \citet{morata05a} combined both single-dish and array
observations, and the resulting maps which represented emissions from both
background and clumped gas confirmed the general picture of \citet{morata03a}.

This small-scale structure had, in fact, been inferred a decade earlier 
from observations of (particularly) CS and NH$_3$ lines in a number of 
objects \citep{myers91a,morata97a}; these studies showed 
unexpected distributions in the emissions from these molecules. 
\citet{taylor96a} proposed that the unexpected behaviour could be 
understood on the basis of time-dependent chemistry in unresolved structures 
that grow and decay on time-scales of around a million years. Alternative 
explanations were also offered (see section 3.2),
but the interpretation of \citet{taylor96a} now 
appears to be supported by the high-resolution observations of - so far - 
two molecular clouds. However, the general picture of clumpiness indicated 
by \citet{peng98a} and by \citet{morata03a,morata05a} is also supported
by discussions of the chemistry associated with interstellar clumpiness 
\citep{cecchi00a} and extending to very small and very 
transient structures in diffuse interstellar clouds \citep{hartquist03a}.

\citet[][hereafter GWHRV]{garrod05a} modelled the chemistry of a single 
transient core within a cloud. They assumed that the clump grows from a 
low density background gas to high density and then decays into the 
background according to a prescription based on a study by \citet{falle02a} 
of the passage of MHD waves through a partly ionized  
cloud \citep[other authors have also found density inhomogeneities may
be produced in low-beta plasma conditions, based on turbulence simulations, 
see e.g.][and references therein]{elmegreen99a}. 
In this model, the chemistry follows the changing density as a 
function of position within the core; ices form on dust in the denser 
parts of the core and are returned to the gas phase when the density 
decreases. The main conclusions of GWHRV were that the dynamical 
time-dependence and the gas grain interactions greatly affected the 
chemistry as compared to that in more conventional models based on static 
objects.

In the present paper we adopt the picture of molecular clouds established
observationally for TMC-1 Core D and for L673; i.e. we assume that a
molecular cloud may be considered as an ensemble of transient cores. 
The observations indicate that a range of core masses is present. For 
simplicity, however, we assume in the present work that all cores have the same 
mass; alternative pictures will be investigated in later work. We
use the chemical results of GWHRV for a single core, and create synthetic
maps at both low and high resolution of an idealized molecular cloud
composed of a large number of such transient cores. These cores are
positioned randomly within our model cloud, and are taken to be at random
stages of evolution in the cycle from low to high density and back again.
Our aim in creating these maps is to determine whether such a model of a
molecular cloud is better able to represent the observed properties of
molecular clouds than conventional static and uniform models, and to use
these models to throw light on the dynamics, chemistry, and gas-grain
interactions occurring in these clouds. Note that we do not attempt here
to model a specific cloud in this work.

In Section 2 we describe the chemical model of a single core used in this 
work, the method of map generation 
and the map parameters adopted. In Section 3 we present results in 
the form of high and low resolution contour maps of emissions from various 
molecules for our model cloud. These show remarkable similarity to 
observational maps. In Sections 4 and 5 we consider the 
stability of our conclusions with respect to changes in map parameters and 
to the adopted mantle re-injection properties. Our general conclusions are 
presented in Section 6.

\section{The Model and Assumptions}
\subsection{The Chemical Model}

We assume that the synthetic dark cloud region is made up entirely of cores
with the chemical and physical parameters of a single generic standard core,
similar to that of the standard run of GWHRV. The new 
standard core differs slightly from GWHRV in several ways:

\begin{itemize}
\item The number of species
is increased by 30, to 251. The new chemical species consist mainly of
longer carbon chain molecules (up to five carbon atoms compared to three).

\item The level of freeze-out is parametrised in a different way; rather
than setting $fr$, the freeze out {\em rate}, to a standard value, it is set
such that it will produce a standard percentage of CO freeze-out for the
central depth point at the time when peak density is achieved.
This level is chosen to be $60\%$. We therefore identify 
$60\%$ CO freeze-out with a sticking coefficient of unity.

\item We choose a threshold visual extinction for freeze-out to proceed of
A$_{V,crit} = 2.5$. The original value of 2 was set to ensure that freeze-out
would have {\em some} noticeable effect; since this is quite clearly not a
problem, we choose a new value more in keeping with the estimate from
\citet{whittet01a} of between 2 and 3. This has the effect of limiting
freeze-out to depth points 7 - 12 (cf. points 6 - 12 in GWHRV).

\item We choose a peak central visual extinction of exactly 5.0 rather than 
5.12.
\end{itemize}

This model produces the same general features as GWHRV, with
the following differences: the higher level of freeze-out results in slightly
higher grain mantle abundances of CO, and higher gas phase levels
of NH$_3$, CH$_4$ and other less abundant carbon-bearing molecules, at
earlier times. CS shows the same double-peaked structure as before, but 
shows stronger central depletion. By the end of the run, any differences
have largely disappeared.

Previous modelling runs have shown that the chemistry of the core is not 
strongly sensitive to a change in evolution time-scale of a factor of two 
shorter or longer, hence our adoption of one core evolution time-scale to 
represent all cores is acceptable.

\begin{figure*}
\plottwo{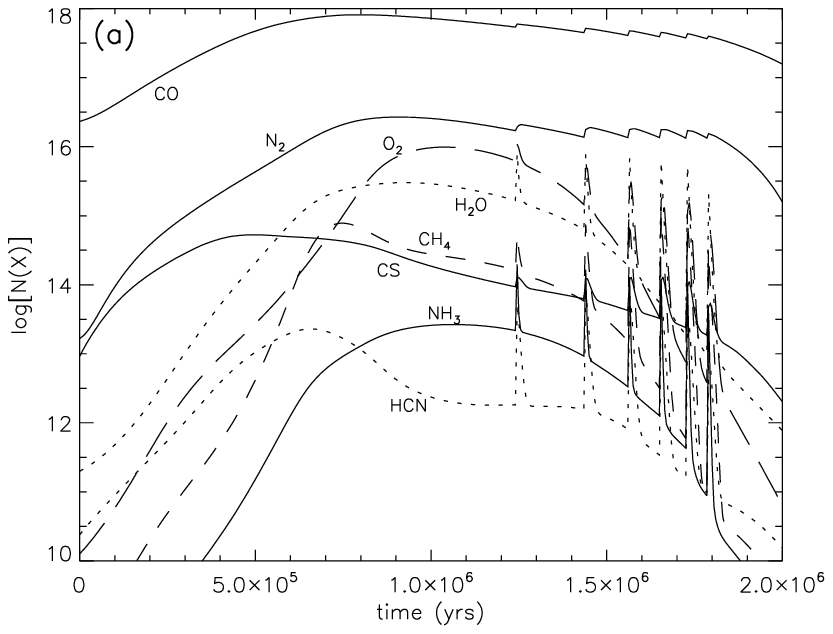}{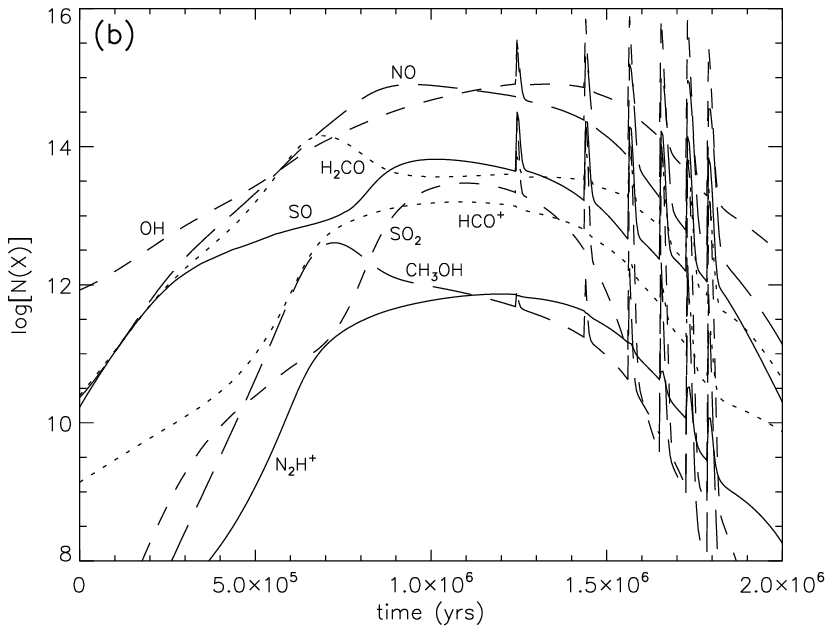}
\caption{\label{fig1} Column densities of selected species as functions of time.}
\end{figure*}

\begin{figure*}
\plottwo{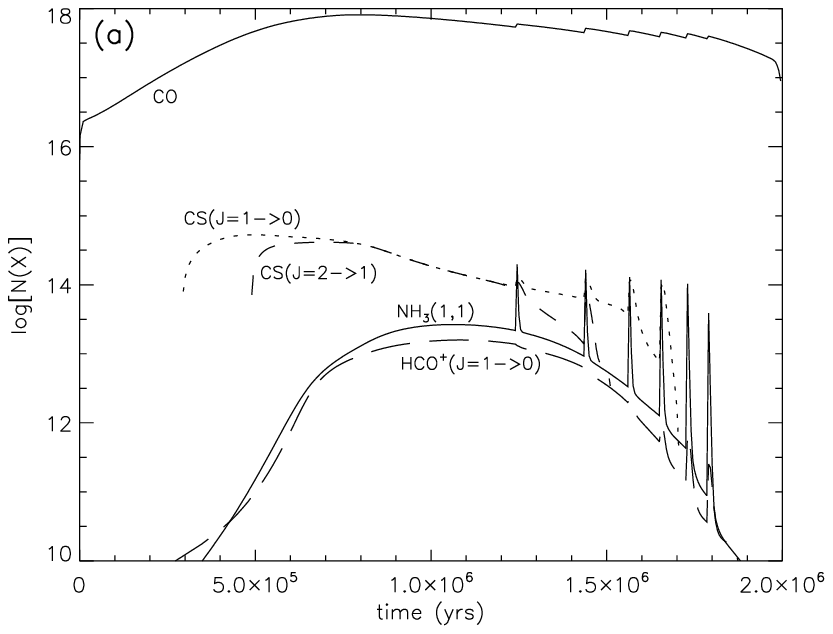}{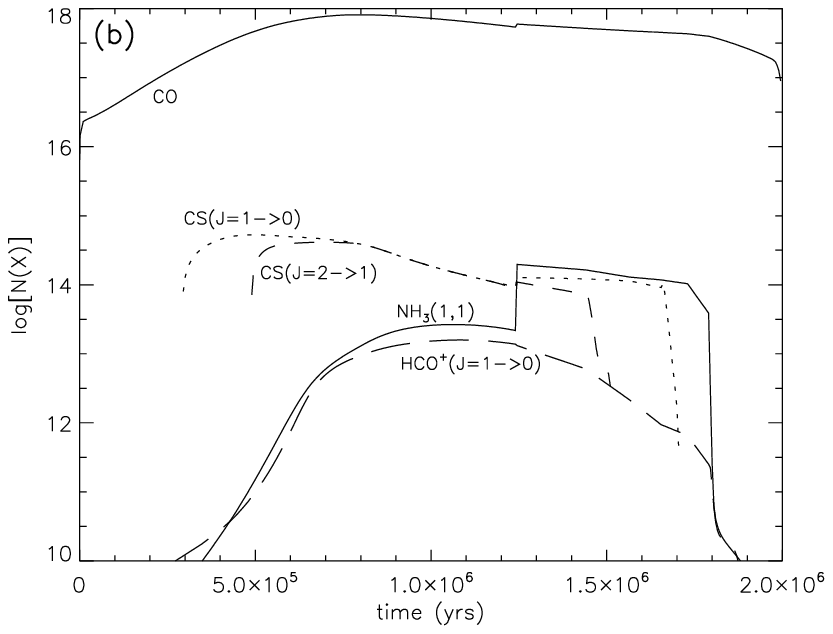}
\caption{\label{fig2} Column densities of selected transitions as functions 
of time, with $n_{\mbox{\scriptsize{{\em eff}}}}$ considerations --
(a) unsmoothed and (b) smoothed assuming maximal re-injection.}
\end{figure*}

An example of the results from the model is shown in figures \ref{fig1}a \& b. 
The discontinuities in the column density profiles after peak time are a
numerical artefact; the
result of the discrete placement of chemical reference points in the core, 
and follow from the instantaneous re-injection of grain ice mantles into the 
gas phase (as discussed in GWHRV). 
We have remedied this numerical problem by integrating the profiles 
between the peaks, effectively eliminating the artificial peak structure
(figures \ref{fig2}a \& b).

This procedure is not entirely consistent with the chemistry, 
since with a continuous distribution of reference points throughout the core, 
we should expect the levels at each point to combine with the others, producing 
larger overall levels. However, the rate of re-injection and the range of 
visual extinctions over which it may take place is not known, since the
mechanism itself is not understood. We therefore consider two extremes:
smoothed column density profiles at the peak values should reasonably 
produce a maximal re-injection peak 
effect. Conversely, by smoothing over the peaks by integrating between the 
levels obtained before the peaks occur and after they have relaxed, we may 
obtain profiles for a minimal level of re-injection. This approximates the case 
where re-injection is slow compared to gas phase reactions. We investigate this
case in section 5.

We make no attempt to model the details of the radiative transfer, and do not 
consider optical depth effects. 
However, we incorporate a very approximate treatment of the excitation: 
We may simplistically argue that for emission in a molecular line to be 
detectable, the gas density must be above the critical density for 
thermalisation \citep[see e.g.][]{evans89a}.
However, as explained in \citet{evans89a}, a more realistic effective critical 
density ($n_{\mbox{\scriptsize{{\em eff}}}}$) which takes into account such 
processes as multi-level transitions and the trapping of line photons in 
optically thick lines, can be defined.
\citet{evans99a} presents values of $n_{\mbox{\scriptsize{{\em eff}}}}$ 
(defined to be the value of the density which yields a line strength of 1K in 
the relevant transition) calculated using a large velocity gradient radiative 
transfer code for some commonly observed molecular transitions, and adopting a 
kinetic temperature of $T_{\mbox{\scriptsize{\scriptsize{K}}}} = 10$ K 
(see Table 1).

We adopt these threshold values and employ them in the calculation of column 
densities using the value of 
$n_{\mbox{\scriptsize{{\em eff}}}}$ as an integration edge, so that material in 
the core for which 
$n_{\mbox{\scriptsize{H}}} \geq n_{\mbox{\scriptsize{{\em eff}}}}$ contributes 
to the emission, whilst material with 
$n_{\mbox{\scriptsize{H}}} < n_{\mbox{\scriptsize{{\em 
eff}}}}$ does not. This approach is still quite simplistic, but acts 
to limit contributions from highly extended regions of gas which may not be of 
sufficient density to emit. This will mostly affect CS morphologies in the 
synthetic maps, since the CS (J=2$\rightarrow$1) line, used in 
\citet{morata03a}, has a high value of $n_{\mbox{\scriptsize{{\em eff}}}}$. CS
is abundant in relatively low density cores at early/late times; the effect
should heavily subdue contributions from these cores in 
the CS (J=2$\rightarrow$1) line, and should cut down the calculated levels even 
in cores at peak densities. 

The critical densities and other information for relevant transitions 
are shown in Table \ref{tab6-1} \citep{evans99a}. Note that data for 
N$_2$H$^+$ were not available. For CO, we assume that a gas density of $1000$ 
cm$^{-3}$ is sufficient to produce a detection \citep{ungerechts97a}, since 
the small dipole moment of the CO molecule produces lines that are usually very 
well thermalised.

Figure \ref{fig2}a shows the unsmoothed column density profiles obtained 
using the effective critical densities of Table \ref{tab6-1}. Figure 
\ref{fig2}b shows the smoothed profiles. Note that CS and NH$_3$ are by far 
the most strongly affected by re-injection. The CS 
(J=2$\rightarrow$1) transition may be seen to trace the cores at closer to the 
peak density than the CS (J=1$\rightarrow$0) transition, although the column 
densities produced are a little lower. The figures show clearly at what stage 
in a core's evolution the densities are too low to produce detectable emission
in the CS transitions.

\subsection{The morphological model}

Having established a chemical model for an individual core, we subsequently
assume that each core has identical chemical and physical/dynamical parameters. 
The only free parameters are the location and evolutionary status of the cores.
To each core we therefore assign column densities calculated for all the 
molecules modelled in the standard run, which vary depending on the designated 
stage of evolution of each core.
To construct a map of the ensemble of cores we adopt the following approach:

\begin{itemize}
\item{Designate a map size, and the number of cores within it}
\item{Assign each core a randomly determined position}
\item{Assign each core a randomly determined stage of evolution}
\end{itemize}
Note that there is no empirical constraint, or model significance, concerning 
the location of the cores in the third dimension.

This results in a purely random ensemble of cores, with no inherent bias
towards any particular structure other than that the region is composed
of discrete cores of gas, and that a uniform core number density has been
adopted for the whole region. This is consistent with the apparent lack of
any particular bias in the L673 region. 
There will, however, be an effect on the resultant local morphologies displayed
within the maps from groupings resulting from the random distribution of the
cores. This will result in unique morphologies for any given distribution of 
cores. 

We also make the important assumption that the cores are chemically and 
physically non-interacting. This is also consistent with the observations of 
individual cores and the findings of the theoretical models. In any 
case, those cores which are localised in the plane of the map may not in fact 
be close neighbours in three dimensions.

We then convolve the integrated column densities 
of each core into a set of molecular line maps for various tracers:
\begin{itemize}
\item{\it Divide map into a grid of ``detection points''.}\\
The resolution of the grid of reference points must be smaller than the beam 
resolution so that the entire map is well covered by the simulated beam.
\item{\it Designate a beam resolution/beam width.}\\
A beam resolution is chosen which simulates a real telescope. A simple Gaussian 
diffraction profile is assumed for the beam.
\item{\it For each detection point, calculate contribution of ``detected column 
density'' from each core which lies within the detection radius.}\\
In this way, simulated detection levels are attributed to each detection point 
in the map from which contour maps can be generated. 
\end{itemize}

To characterize the size of the cores, we adopt a Gaussian profile, and define 
a scale width as the intrinsic radius. Since the models of \citet{falle02a} are 
only one-dimensional, we cannot say with any certainty how 
the density profiles along other axes may vary. However, the adoption of a 
Gaussian profile is in line with the density profile approximations along the 
axis of core collapse. We make one further 
constraint on the intrinsic radius of the core: that it obeys the conservation 
of mass requirement (GWHRV). The \citet{falle02a} mechanism does not 
necessarily require the cores to conserve their mass, however this stipulation 
is in line with our assertion that the cores are individual, non-interacting 
entities. Because all of the column density calculations obtained from the 
chemical model are made along the principal collapse axis, for the sake of 
consistency we assume that each core is aligned with its principal collapse 
axis along the line of sight. Therefore, we allow the observed radius of each 
core to vary only in a way consistent with this alignment, that is:
\[
\Delta r(t) \propto [\Delta z(t)]^{1/4}
\]
where $\Delta z(t)$ is the core scale width along the principal axis as defined 
in GWHRV (see Appendix for derivation) and is calculated in the 
chemical model. 
$\Delta r(t)$ obviously reaches a minimum at the time of the peak density of 
the core, $t_m$. 
We fix $\Delta r(t_{m})$ so that it assumes a value of some (substantial) 
fraction of $\Delta z(t_{m})$, determined according to comparisons with 
\citet{morata03a}.

Having attributed each core with an intrinsic width dependent on its stage of 
evolution, the map of the cores of finite dimensions is convolved with the 
(off-axis) Gaussian beam to produce the synthesized maps.
Note that the column densities only account for the centre of the core, 
along the line of sight, so the ``detected'' morphologies should be thought of 
as being representative of the extents of the {\em core centers} rather than 
the entire core plus envelope. 

The uniform alignment of the cores' collapse axes 
along the line of sight is a necessary simplification, given the 1-dimensional 
nature of reliable MHD simulations \citep[see GWHRV;][]{falle02b} on which to 
base our dynamical model. The main effect of this assumption is to make every 
core circular in the plane of ``observation''; we might expect that a full 
3-dimensional chemical/dynamical model would produce non-circular morphologies 
for individual cores, if even higher resolutions than we use here were adopted. 
It should not have a drastic effect on the convolved morpholgies shown in this 
work. We would also expect that the relative extents in the plane for molecules 
in individual cores could be somewhat different (e.g. CS versus NH$_3$; see the 
collapse-axis molecular extents shown in GWHRV). However, this effect should 
not be large since contributions are strongest from the core centers. It could 
be remedied by choosing core widths according to the molecule being mapped, 
however for consistency we use the same widths for all molecules.

\subsection{The Adopted Map Parameters}

There are approximately 40 spatially separable cores observed in the CS 
(J=2$\rightarrow$1) line in the $\sim$(0.5 pc)$^2$ maps of \citet{morata03a}.
The cores observed in the HCO$^+$ (J=1$\rightarrow$0) and N$_2$H$^+$ 
(J=1$\rightarrow$0) lines do not generally coincide with those of CS or with 
each other; this is as expected (GWHRV) and is primarily a result of 
chemical differentiation within the cores. 
Also, some closely separated cores will be unresolved whilst others will have 
observationally undetectable column densities. We therefore estimate a value of 
$\sim$60 cores, leading to a core number density of 250 pc$^{-2}$. 

\begin{figure*}
\epsscale{.60}
\plotone{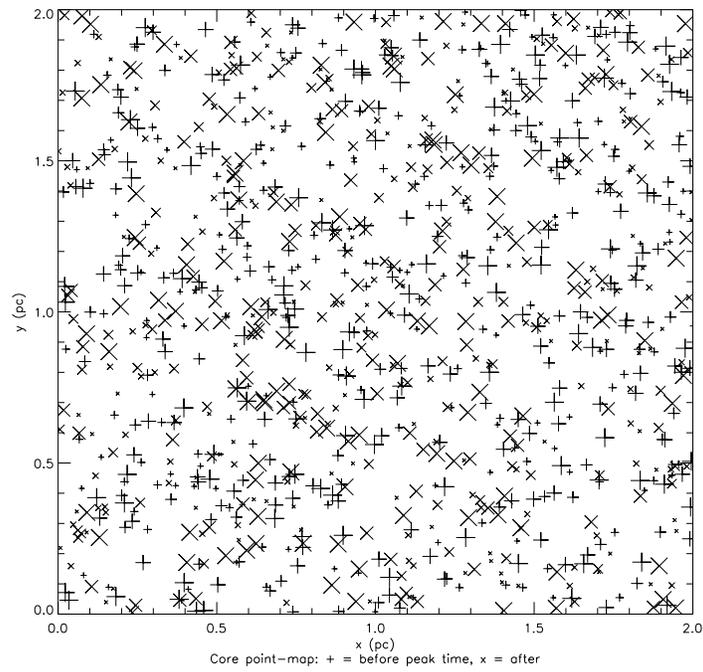}
\caption{\label{fig3} Core positions in the basic ``sky'' map. `+' indicates a 
pre-peak density core, `x' indicates a post-peak density core. Symbol size 
indicates proximity to peak density time, occurring halfway through core 
evolution.}
\end{figure*}

We use a global map size of 2 pc$\times$2 pc  \citep[approximately equal to 
the size of the CS maps of][]{morata97a} within which there are a total of 
1000 cores. Figure \ref{fig3} shows the basic map of core positions, the 
synthetic ``sky''. We denote the approximate 
stage of evolution: Plusses (+) indicate cores that have not yet reached peak 
time, and are therefore not yet at maximal density; crosses ($\times$) indicate 
cores which have evolved past the peak time. The relative sizes of the symbols 
also indicate approximately how far into the cycle the cores are: larger 
symbols indicate larger densities (or proximity to peak time).

We take two subsets of the data in these maps which simulate the observational 
data:

\begin{itemize}
\item {\it Large-scale, low resolution maps}.\\
These maps essentially cover the whole 2pc$\times$2pc area, although we ignore 
``emission'' from within 0.1 pc of the edges.
This is because levels at the edges are influenced by the lack of any 
cores outside of the map, producing an unnatural drop in ``detection''. To 
avoid this artefact, we remove a strip around the edge of the maps of size 
approximately equal to the beam radius. 
The resolution corresponds to a beam width of just under 2 arcmin to agree 
approximately with \citet{morata97a}. This corresponds to a beam radius of 
0.085 pc, calculated at a distance to L673 of 300 pc.

\item {\it Small-scale, high resolution maps}.\\
Here we zoom in on a region of the low resolution maps of size $\sim$ 0.6 pc 
$\times$ 0.6 pc chosen so as to accommodate both local CS and NH$_3$ peaks.
This is comparable to the area ($\sim$0.5 pc $\times$ 0.5 pc) studied by 
\citet{morata03a}. This was chosen from within the detection map of
\citet{morata97a} to coincide with peaks in CS emission, and in NH$_3$ detected 
by \citet{sepulveda01a}. 

The synthesised beams employed in \citet{morata03a} using BIMA are 
$\sim 20 \times 15$ arcsec, corresponding to a circular beam radius used here 
of $\sim$0.0125 pc.
\end{itemize}
Note that a single beam width is adopted for all species/transitions.

\section{Results}

\subsection{Low resolution, large-scale maps}

\begin{figure*}
\epsscale{0.35}
\plotone{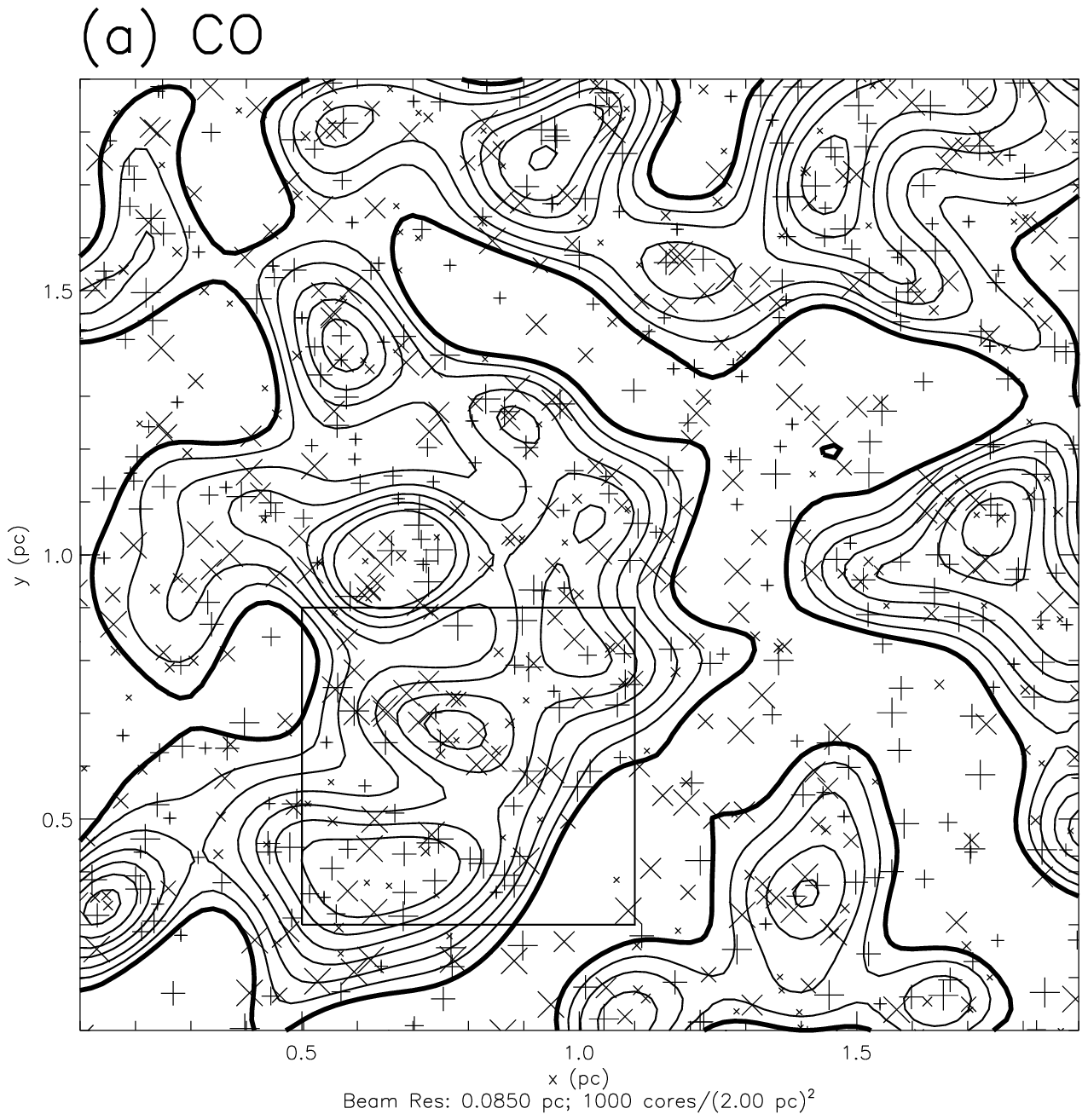}
\plotone{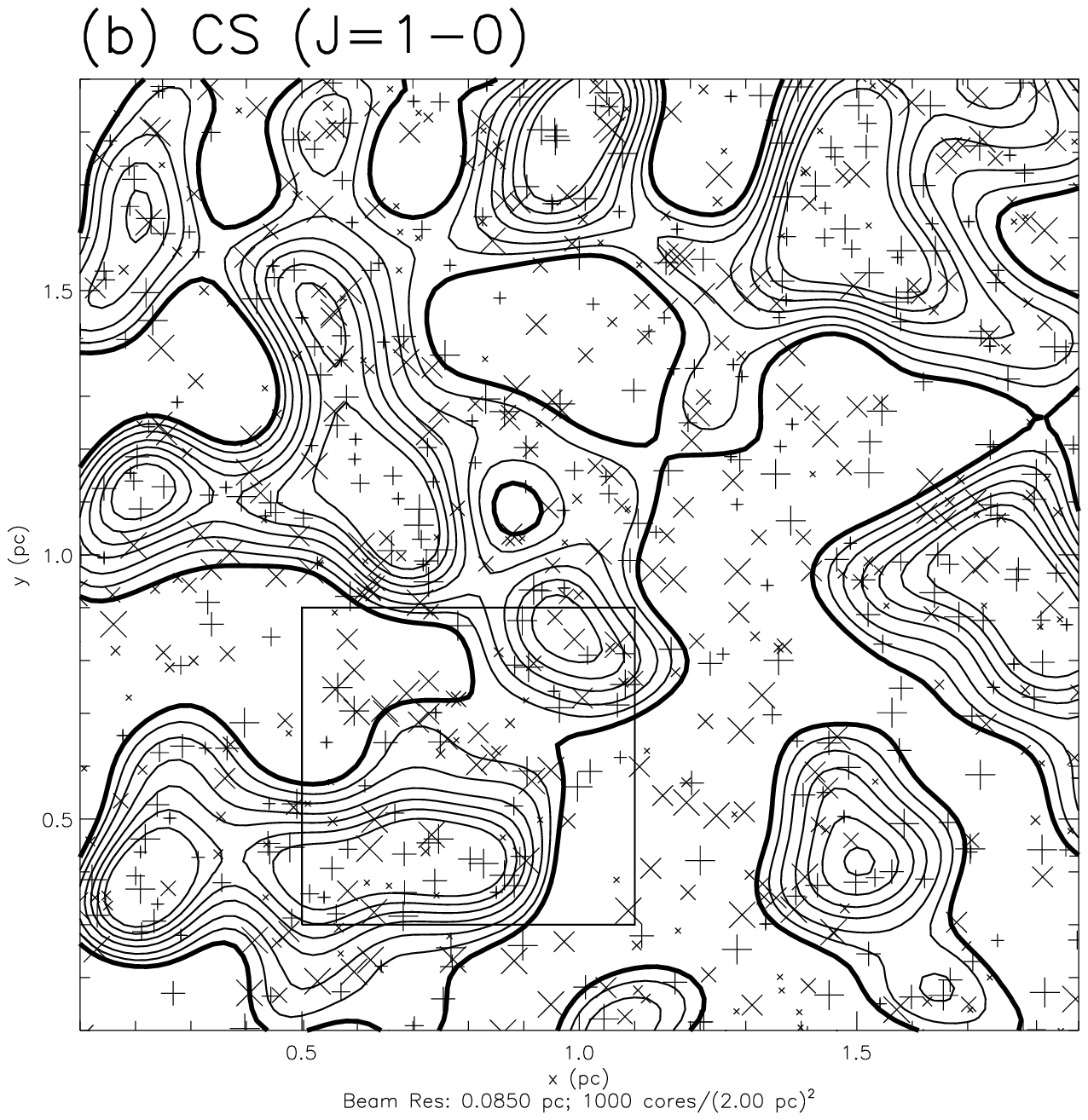}
\plotone{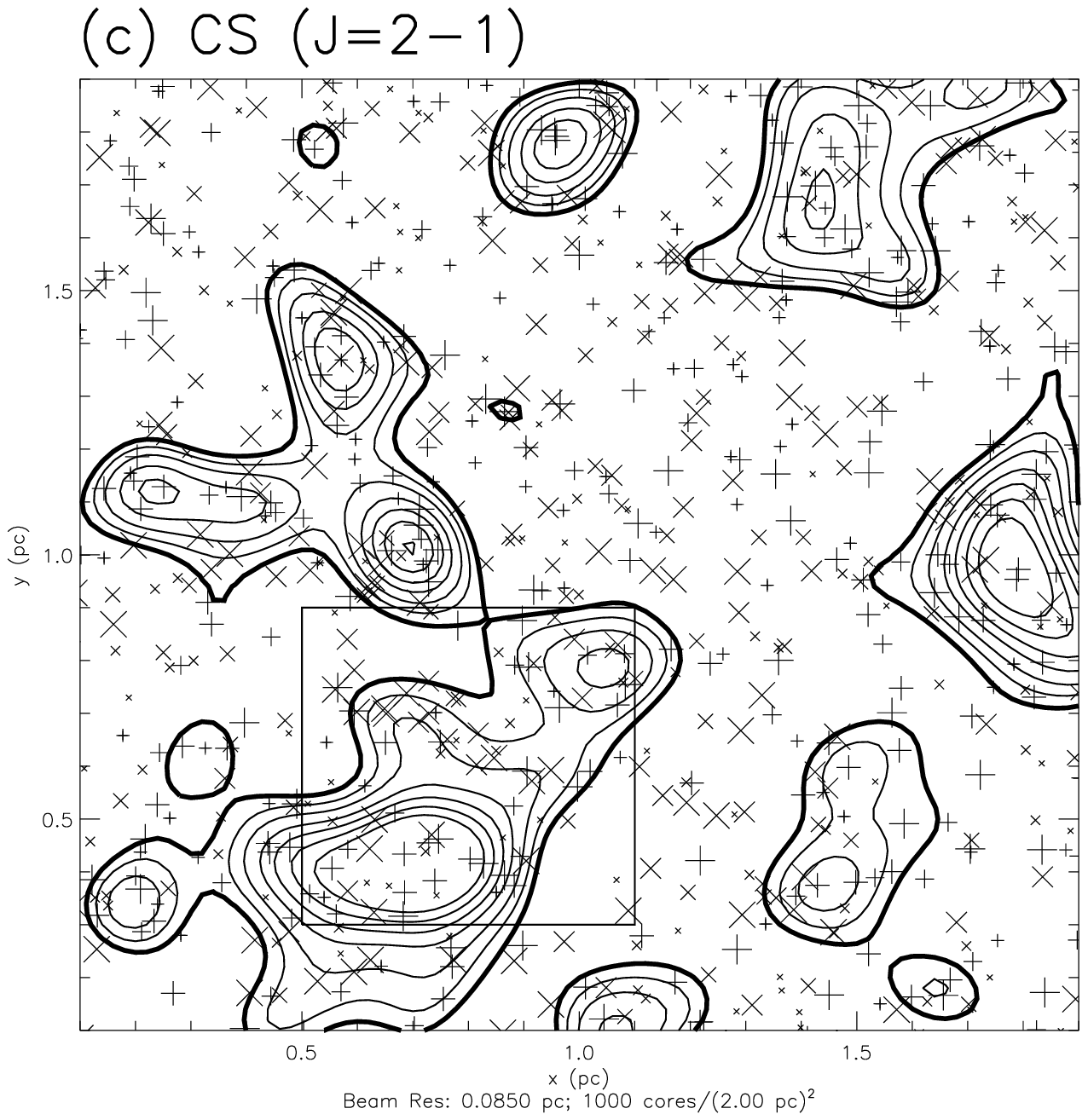}
\plotone{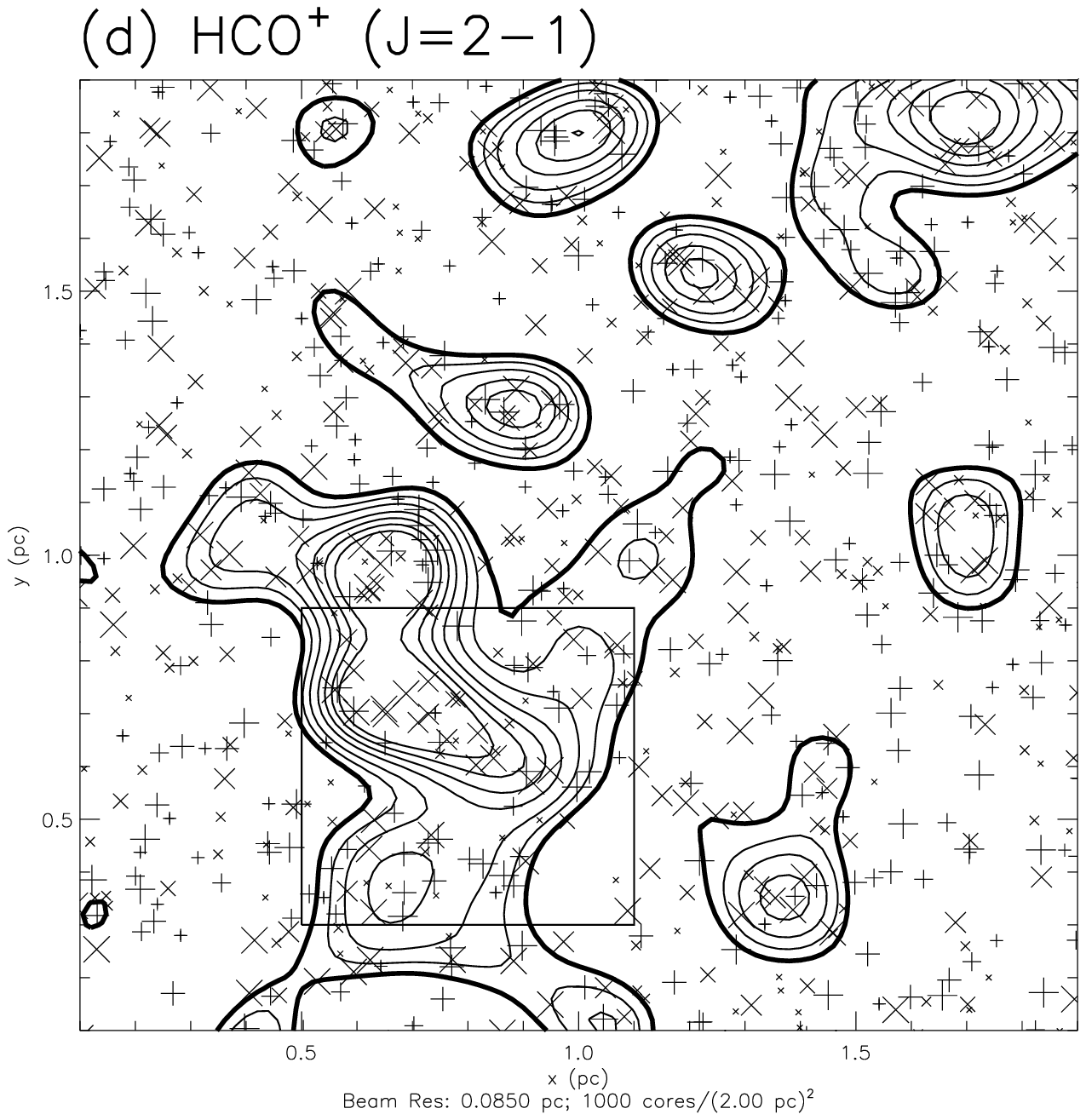}
\plotone{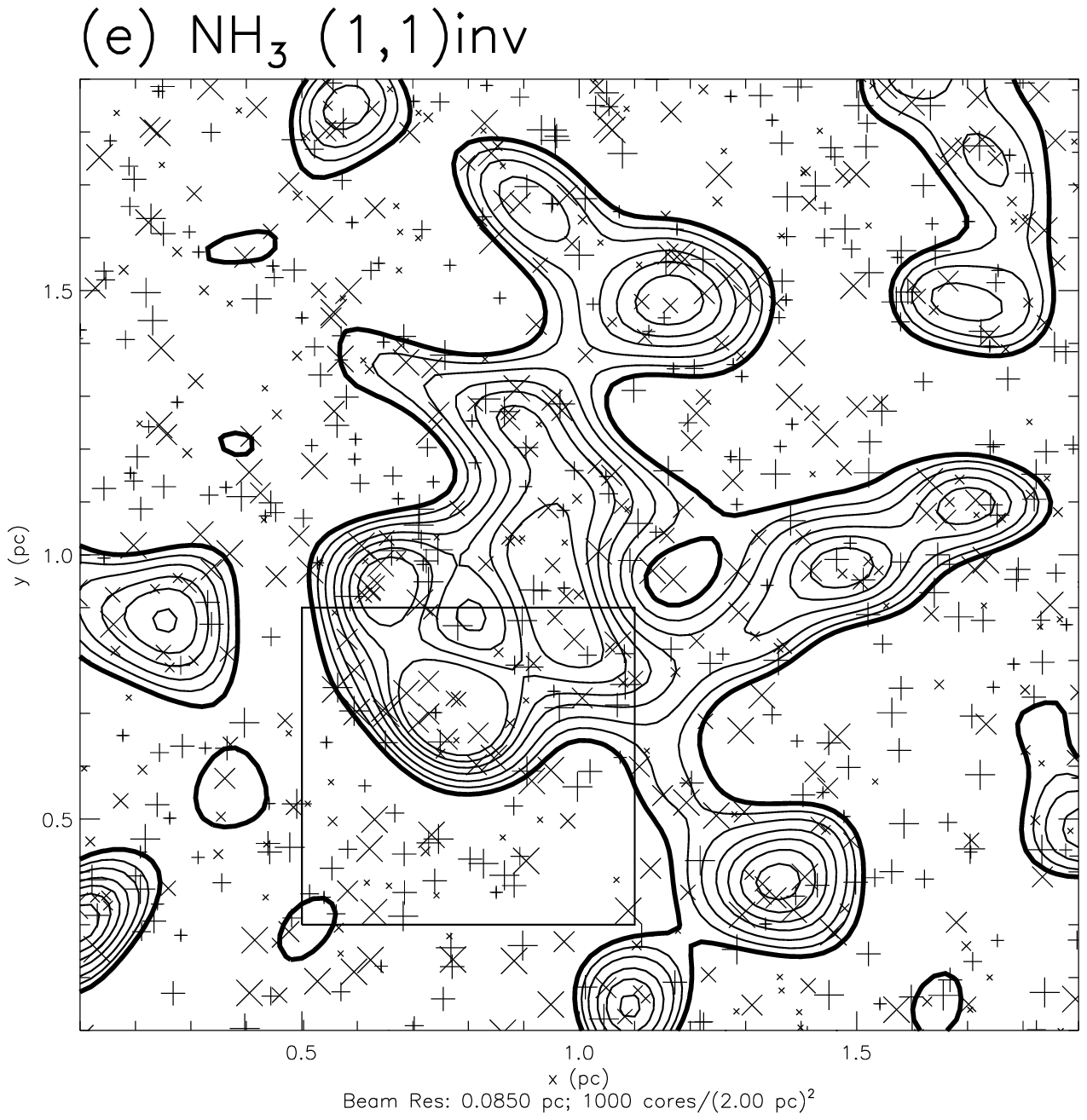}
\plotone{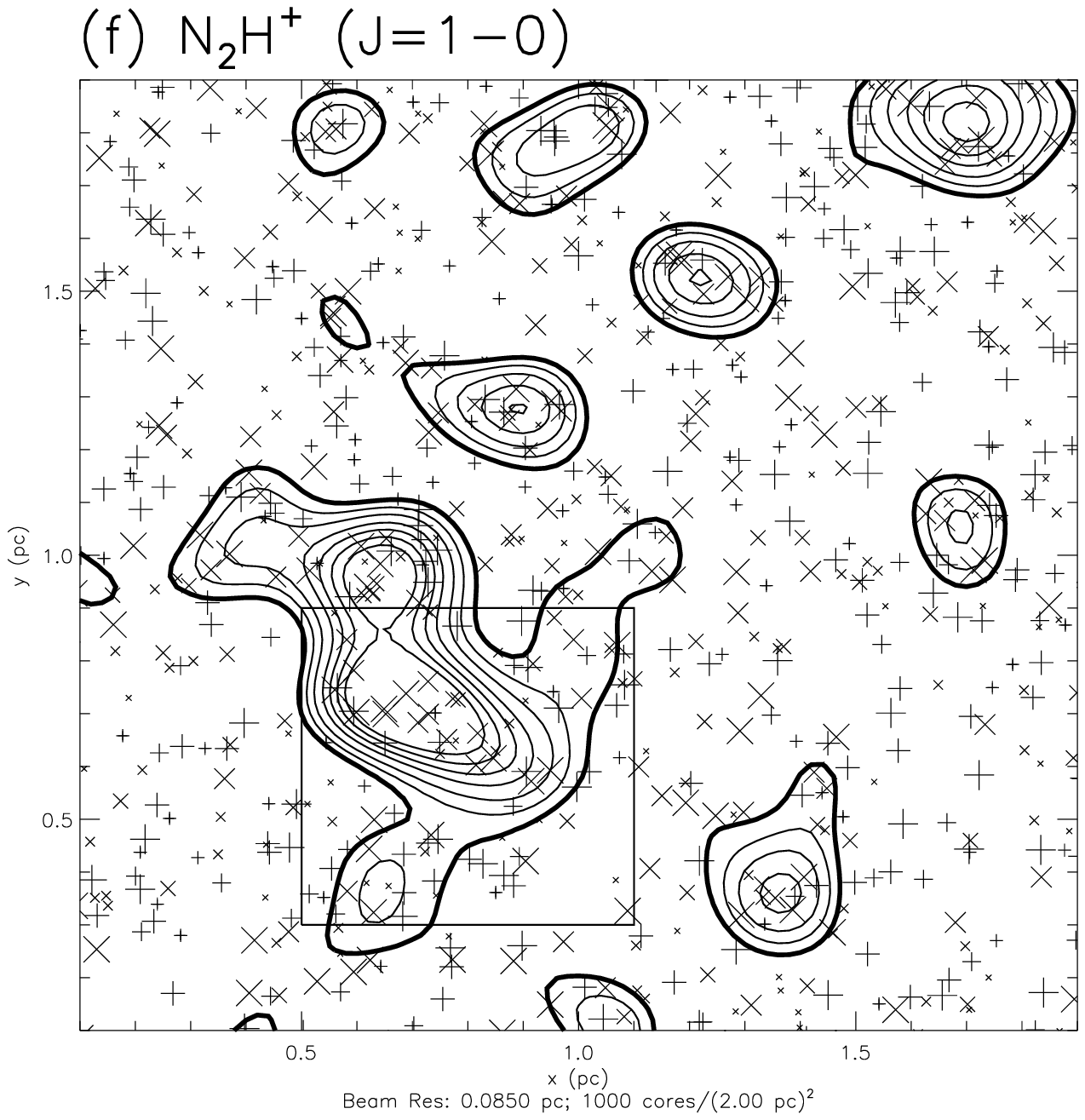}
\caption{\label{fig4} Low resolution convolved maps, beam FWHM$=0.17$ pc. Maps 
are trimmed by approximately 1 beam radius to remove effect of artificial map 
edge in the contour plots. Box in lower left indicates zoomed region shown in 
figure 5.}
\end{figure*}

Figures \ref{fig4}a -- f show low resolution ($\sim$2' FWHM) {\em column 
density} convolved molecular line maps for six density tracers. We 
overlay the basic map of core positions (Fig. \ref{fig3}) for ease of
comparison. The contours represent fractions of the peak value calculated for
the map, for any particular 
molecule. The fractions of peak ``emission'' represented by the contours are 
chosen to correspond with those of the low resolution observations shown in 
\citet{morata03a}. The outermost (thick) contour line represents the 
half-maximum (0.5), with the levels rising by $\sim$0.05. The overlaid boxes 
represent the portion of the map on which we zoom-in in the high resolution 
maps.

As is generally observed in dark clouds \citep[see e.g.][]{myers91a} the 
morphologies of the different molecular tracers are not the same, do not trace
the same regions, and have different spatial extents. 
Indeed we see here that the characteristic features exhibited in molecular 
line maps of CO, CS and NH$_3$ in \citet{myers91a} are broadly re-produced. 
We find that CO and CS are not significantly different in spatial extent.
The ``detection'' of NH$_3$ is comparatively compact and agrees well with the
evidence of \citet{myers91a}.
HCO$^+$ and N$_2$H$^+$ both seem to track approximately the same regions of 
gas, with very similar morphologies and extents. These molecules display 
smaller extents than CO or CS and their contour map peaks correspond to peaks 
in H$_2$ and CO. This is consistent with the observation that N$_2$H$^+$ 
tends to be a good density tracer.\\

\subsection{High resolution, small-scale maps}

For the zoomed-in high resolution synthetic molecular line maps, we choose a 
region from the bottom-left section of the map of approximately the same size 
as the region of L673 examined in \citet{morata03a}, chosen for the presence of 
CS and NH$_3$ peaks in that area.

\begin{figure*}
\epsscale{0.35}
\plotone{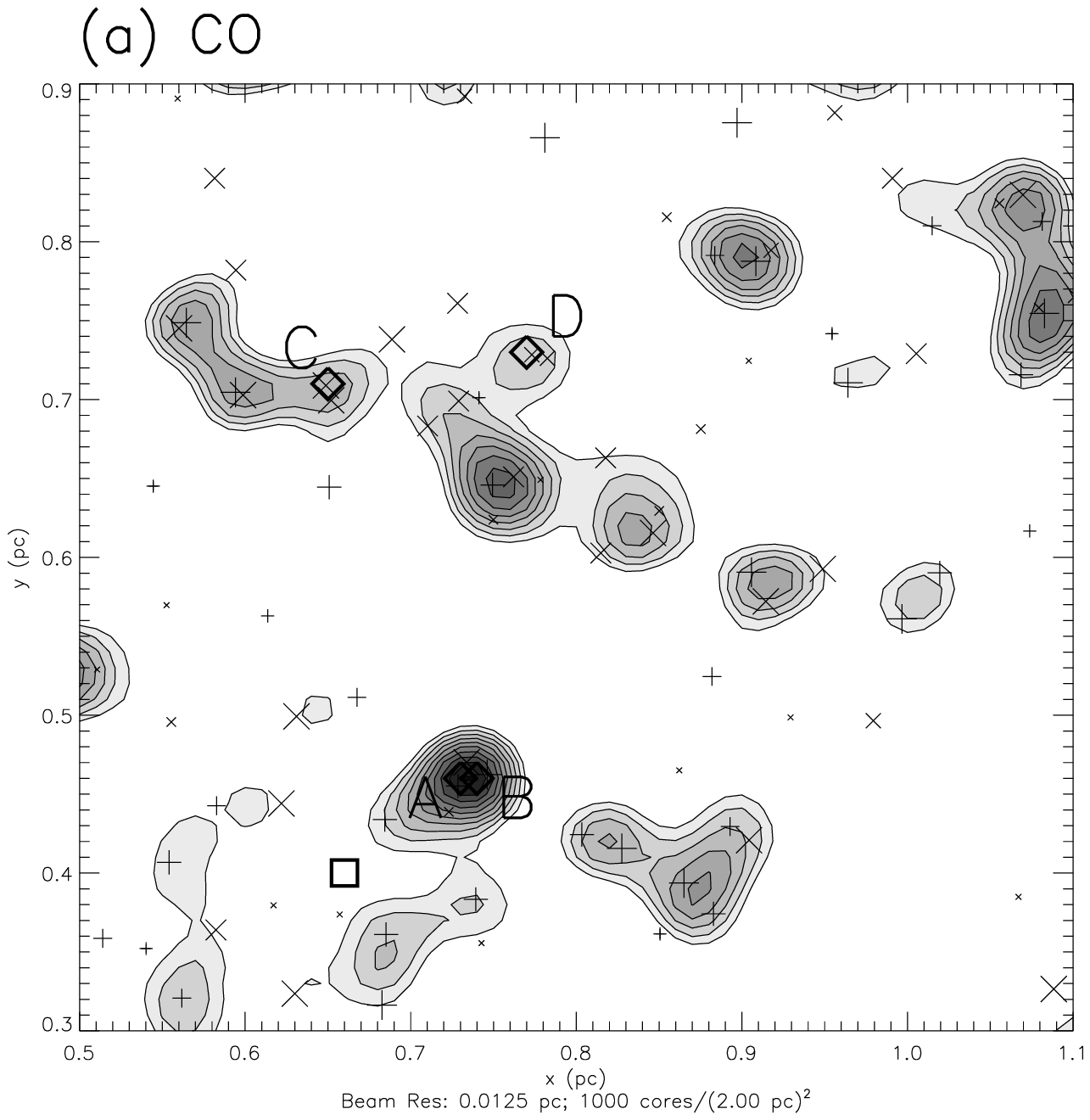}
\plotone{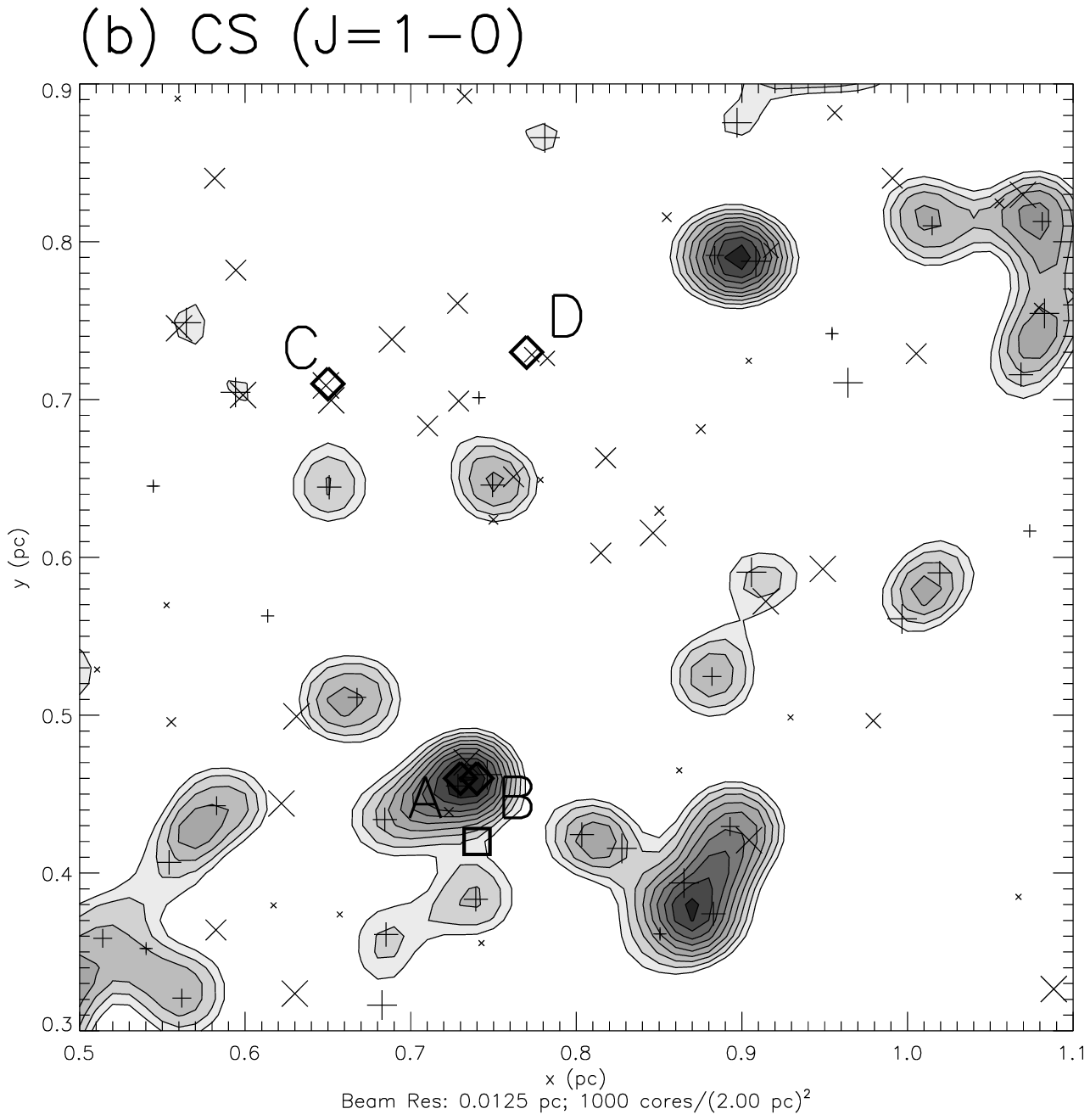}
\plotone{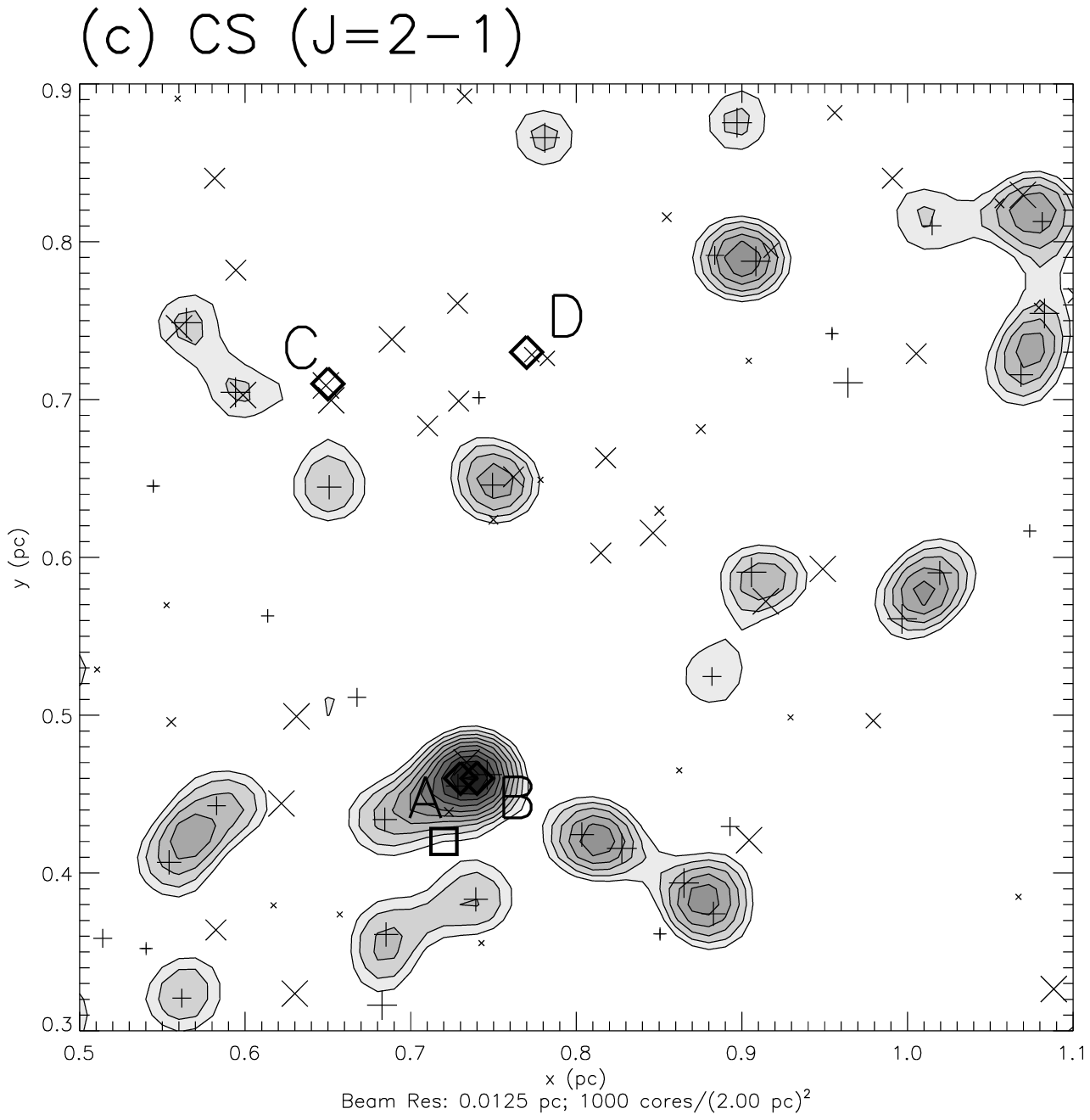}
\plotone{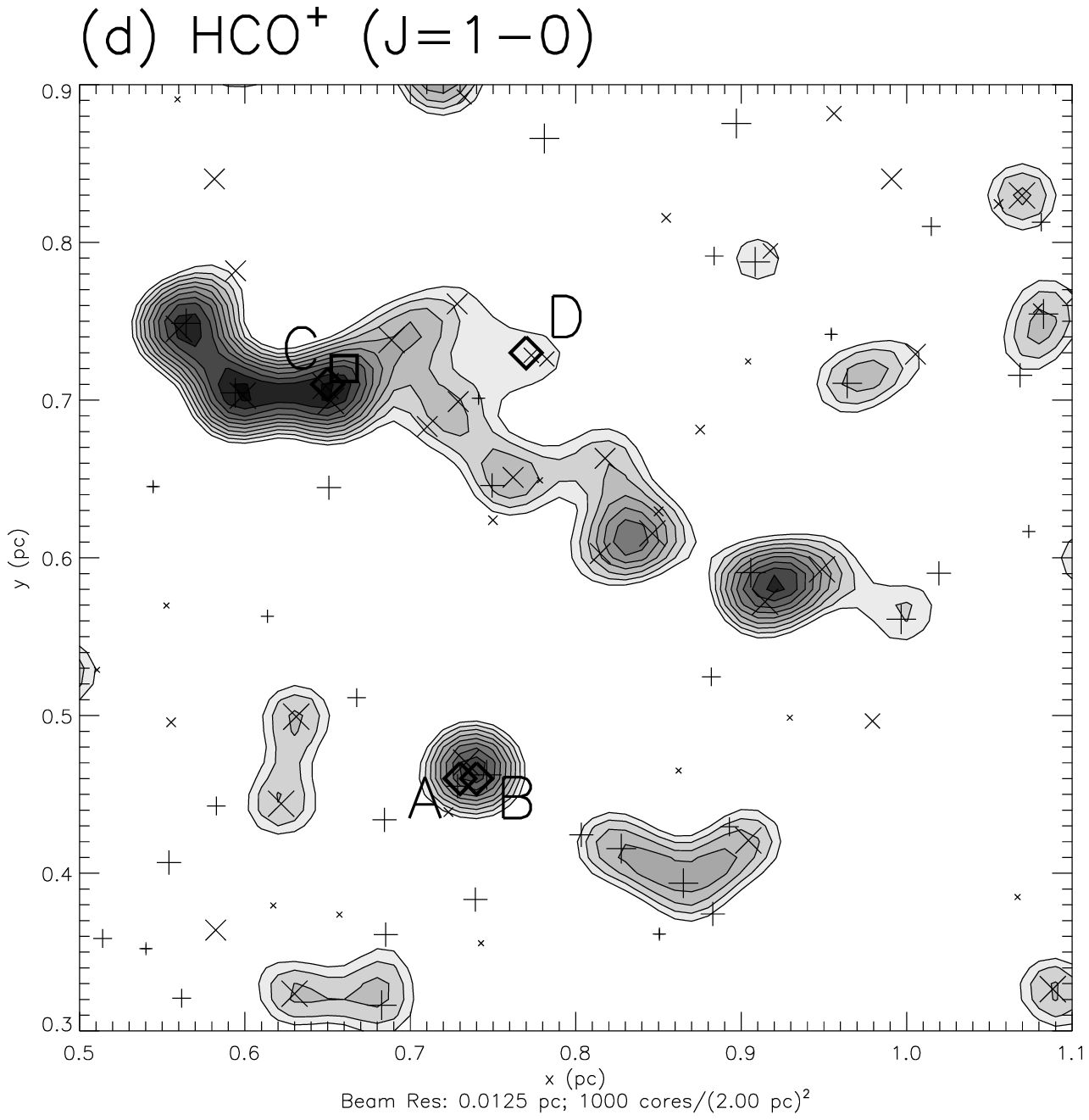}
\plotone{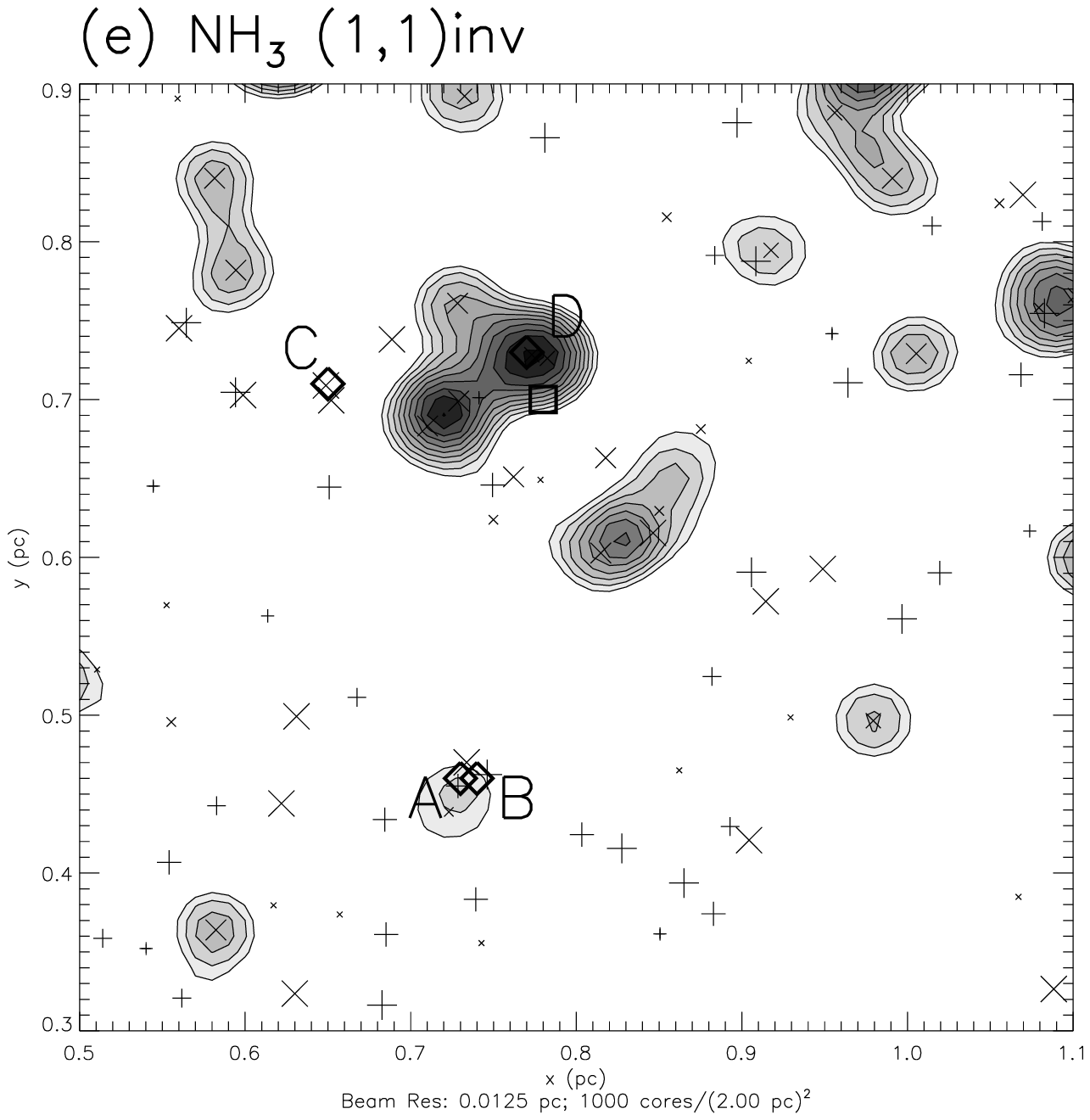}
\plotone{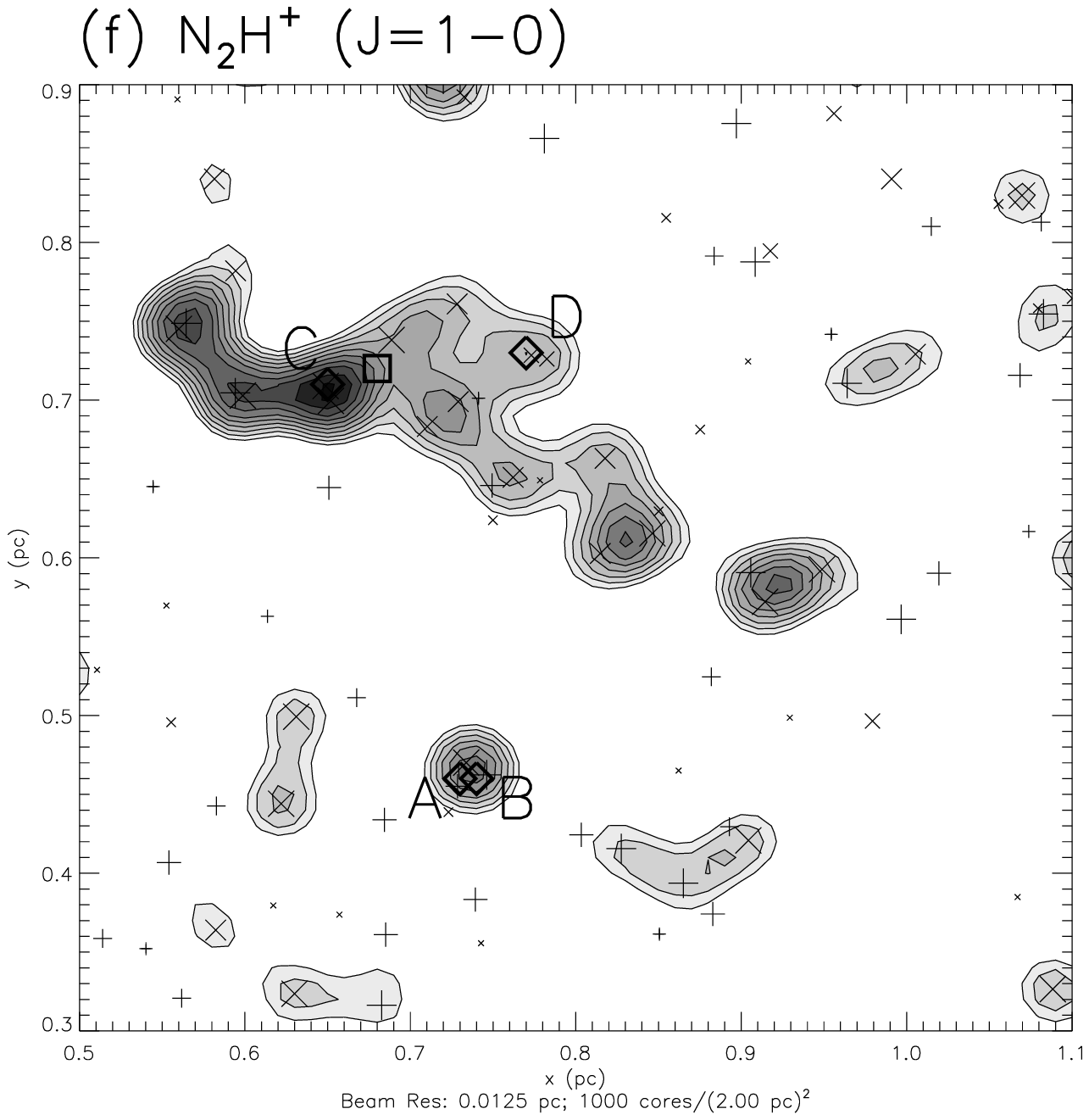}
\caption{\label{fig5} High resolution zoomed maps, beam FWHM$=0.025$ pc. Map 
region corresponds to boxed area in figure 4. Square symbol indicates low 
resolution peak of mapped molecule. Diamond symbols correspond to high 
resolution peaks in: (A) CO, CS (J=1$\rightarrow$0); (B) CS 
(J=2$\rightarrow$1); (C) HCO$^+$ (J=1$\rightarrow$0); (D) NH$_3$ (1,1)inv.}
\end{figure*}

Figures \ref{fig5}a -- f show maps for CO, CS, HCO$^+$, NH$_3$ and 
N$_2$H$^+$. Contour levels are set to correspond to 
those used in \citet{morata03a}: the lowest is 0.36 of maximum, the others 
increasing by 0.07.

In comparison with \citet{morata03a}, the CS morphologies are generally 
similar, although we do not see so many resolved cores in the synthetic map; 
the same is true for HCO$^+$. Although the overall morphology for N$_2$H$^+$ 
is very similar to that of HCO$^+$, the number of resolved N$_2$H$^+$ cores 
matches well with \citet{morata03a}. Both the larger- and smaller-sized cores 
are reproduced in the synthetic maps. Note that the ``emission'' is not purely 
centred around individual cores but around small groupings. Clumps of emission 
out on their own which reach only two or three contour levels (0.43 -- 0.50 
of maximum) tend to be the result of individual cores or loose associations of 
two or three.

The good agreement of comparative CS and NH$_3$ peak positions and extents in 
the low resolution maps with the observational evidence of 
\citet{morata97a} is encouraging: The general 
morphologies of these molecules match those of dark cloud cores examined in 
\citet{myers91a}. Those authors calculated mean FWHM extents in 16 dark cloud 
cores to be 0.15 pc (NH$_3$), 0.27 pc (CS) and 0.36 pc (C$^{18}$O). The extents 
of CS and NH$_3$ are arguably larger than this, but appear to preserve the
correct relative extents. 
This is particularly important in that the spatial extents of the CS and NH$_3$
emissions are the opposite of what one would expect from simple critical
density arguments {\em for a single core} \citep[see e.g.][]{rawlings96a}.
Previous attempts to resolve this problem \citep[e.g.][]{tafalla02a} have 
invoked differential depletion effects between sulphur-bearing and 
nitrogen-bearing species \citep[following][]{bergin97a}, but the chemical 
basis for such models has not been confirmed. 

\citet{taylor96a} first suggested that the CS/NH$_3$ 
discrepancy could be a result of the clumpy nature of molecular clouds. Their 
chemical models used the modified free-fall collapse mechanism of 
\citet{rawlings92a}, and did not include subsequent dispersal. They 
conjectured that most clumps would disperse before significant NH$_3$ levels 
could build up, whilst CS should peak before this time arrived. The adoption 
of the mechanism proposed by \citet{falle02a}, here, allows for the 
(controlled) dispersal of a core. However it is for similar reasons as those 
suggested by \citet{taylor96a} that the differences in the morphologies of CS 
\& NH$_3$ is reproduced in our models; the NH$_3$ peaks later than CS, and 
for a less extended period of time.

The good morphological agreement of the high resolution maps, especially for 
N$_2$H$^+$, with those of \citet{morata03a} indicates that the \citet{falle02a} 
mechanism produces data consistent with observations of small, transient, dense 
cores. Whilst the simulated low resolution maps should not be expected to 
correspond closely with the particular example of region L673, we note that, as 
in L673, the emission from CS and HCO$^+$/N$_2$H$^+$ do not generally represent
the same cores, regions of stronger emission overlap, and similar 
ratios of strongly and weakly emitting cores are present. 

In Table \ref{tab6-2} we present column densities (and ratios) as determined 
by the model, 
for the low resolution and high resolution maps, and compare with levels 
detected in \citet{morata97a} and \citet{morata03a}. Taking a similar approach 
to the latter work, we select the peak position for each transition in the 
high resolution maps and compare values at the same spatial positions. 
These points are represented by {\em diamond}-shaped markers in figures 
\ref{fig5}b--e. We also give column density values calculated from the low 
resolution data, at the low resolution molecular line maxima within the high 
resolution map domain. These are represented by {\em square}-shaped markers. 
Following \citet{morata97a}, for each low resolution peak we present the column 
density only of the molecule whose peak it is. Peak A 
corresponds to the shared high resolution peak in CO and CS 
(J=1$\rightarrow$0). Peaks B, C, and D correspond to the high resolution peaks 
in CS (J=2$\rightarrow$1), HCO$^+$ (J=1$\rightarrow$0) and NH$_3$ (1,1), 
respectively.

In Table \ref{tab6-3} we present the equivalent column density data obtained 
from the low resolution observations of \citet{morata97a} (their table 4) and 
the high resolution observations of \citet{morata03a} (their table 3). Peaks 
E and W correspond to CS (J=2$\rightarrow$1) peaks, whilst peak N corresponds
to HCO$^+$ (J=1$\rightarrow$0) and peak S to N$_2$H$^+$ (J=1$\rightarrow$0).

Considering the low resolution peaks in Table \ref{tab6-2}; the CS 
(J=2$\rightarrow$1) maximum (B) is close in value to the CS 
(J=1$\rightarrow$0) peak level (A) - and the two peaks are located very close 
together. The low resolution HCO$^+$ value is more than an order of magnitude 
lower than the peak in either of the CS transitions. NH$_3$ is a factor of 
2 -- 3 lower than 
the CS peak levels. The distance between the NH$_3$ and CS (J=1$\rightarrow$0) 
low resolution peaks is 0.31 pc, which agrees reasonably well with the 
$\sim0.2$ pc quoted for the specific case of L673 in \citet{morata97a}. 
However, the actual 
levels and ratios detected do not agree well with that paper. Whilst the 
synthesised NH$_3$ peak level may be as low as half the observed value, the 
synthesised CS (J=1$\rightarrow$0) peak level is about an order of magnitude 
too high. This leaves a ratio of [CS/NH$_3$]=3.0, rather than the quoted value
of $\leq 0.11$.

For the high resolution synthetic values, at the CS (J=2$\rightarrow$1) peak 
(B), the HCO$^+$ (J=1$\rightarrow$0) column density is little changed from its
peak value (C). However, the CS (J=2$\rightarrow$1) column density at peak C 
is a factor of three lower than at B. This difference is due to peak B being 
an unresolved combination of three strongly emitting cores -- one slightly
post-peak density, another slightly pre-peak density, and the other a little
younger again. This combination acts to give high values for each of the
molecules except NH$_3$ (which is so strongly weighted towards later times in
the column density profile of figure \ref{fig2}). The close proximity of the
three cores, which are approximately 1 high resolution FWHM beamwidth apart,
allows them to reinforce each other effectively.

The high resolution synthesised CS (J=2$\rightarrow$1) and HCO$^+$ 
(J=1$\rightarrow$0) column densities are even larger again than the observed 
values -- peak CS levels are as much as two orders of magnitude larger than 
observed, and HCO$^+$ levels more than one order of magnitude larger.
It can be seen that, contrary to the observations, the synthesised column
densities are higher at the high resolution peaks than for the low resolution
peaks. This is possibly an effect of the presence of many, unresolved, 
low-level cores in L673, or else the interferometric observations may have 
resolved out the more extended features.

The overall column density levels obtained are generally too high, even at low
resolution. Four factors may explain this: 
(i) the level of uniformity of core distribution may have an 
effect such that the peaks observed in the synthetic maps represent close 
associations of cores (as mentioned above), in particular those close 
associations of cores at similar stages of evolution,
(ii) the chemical (and/or physical) evolution of individual cores may vary from
that adopted here, 
(iii) this approach uses only one core chemistry to model the entire region. 
In reality, we might expect to see a spectrum of cores within the same cloud. 
Chemical model runs adopting different parameters from those of the standard
core run have shown that the resultant chemistry (and the CS \& NH$_3$
abundances) can vary strongly with visual extinction,
(iv) smaller core widths (in the plane) should produce smaller column 
densities. The \citet{morata03a} HCO$^+$ emission is less extended and more 
defined than the CS in particular, and this may be due to only smaller regions 
near the core centre being able to produce it. 
Evidently, the model cloud is chemically richer than L673. No attempt has been 
made here to model L673, other than in general morphology. It may be, for 
example, that the cosmic ray ionization rate is less for L673 than adopted in 
the model.

\section{Testing the Effects of the Variation of Core Distribution}

As suggested above, the precise morphological distributions depend on the
assumed (random) distribution of the cores. In this section we investigate the 
sensitivity of the results to this distribution.
We restrict the analysis to the high resolution maps so that the assumed 
distribution is locally valid.

We define two statistical measures of the uniformity of the distribution of the 
cores: the mean shortest inter-cores distance (ICD, analogous to core number 
density) and the standard error in this mean (SEM, uniformity of distribution). 
In the locality of a single core, the ICD is very 
important for the relative influences of either core on the convolved map, 
especially when close to the beam resolution limit. Higher values of the SEM 
imply greater deviations from uniformity.

\begin{figure*}
\epsscale{0.35}
\plotone{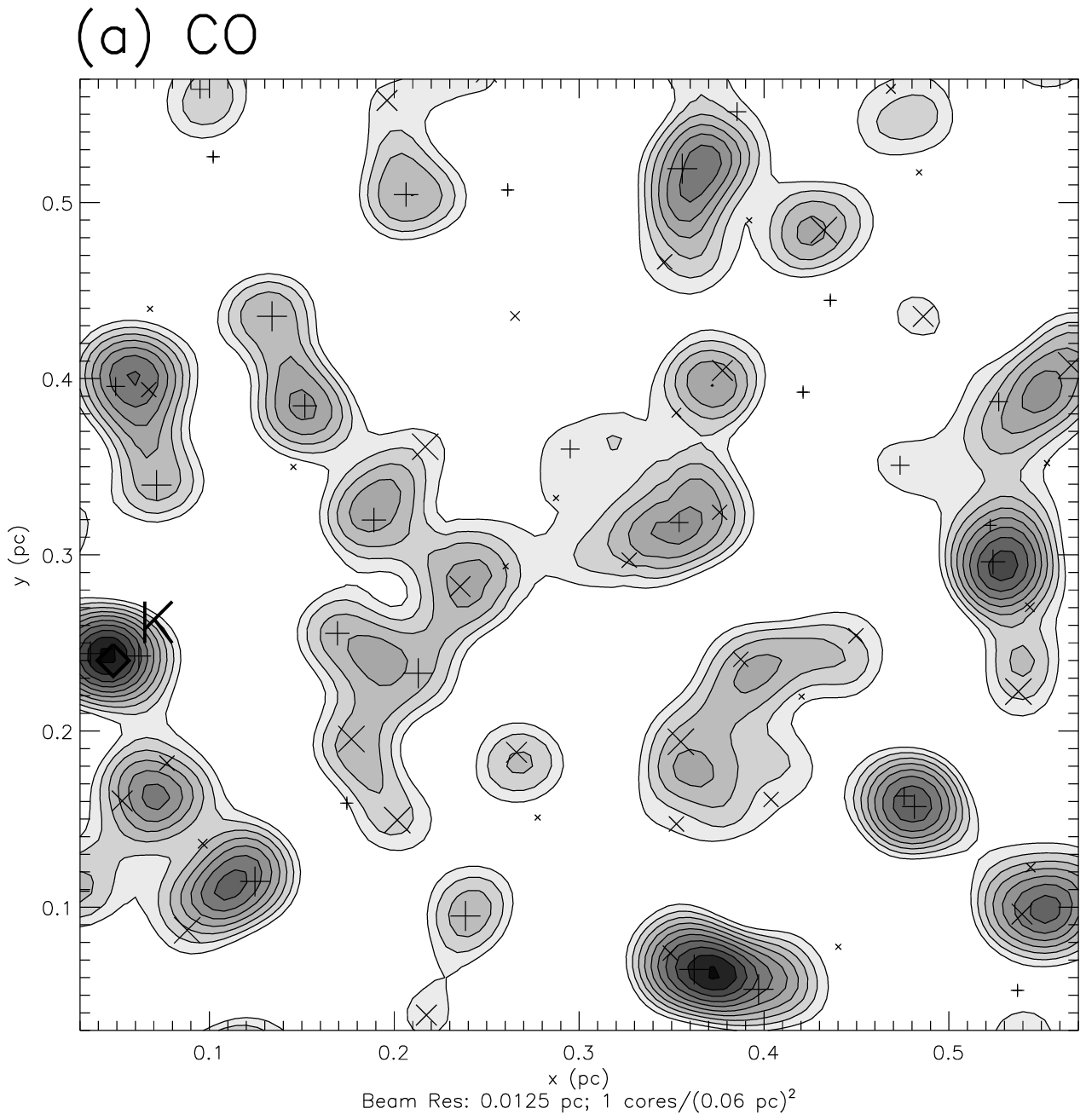}
\plotone{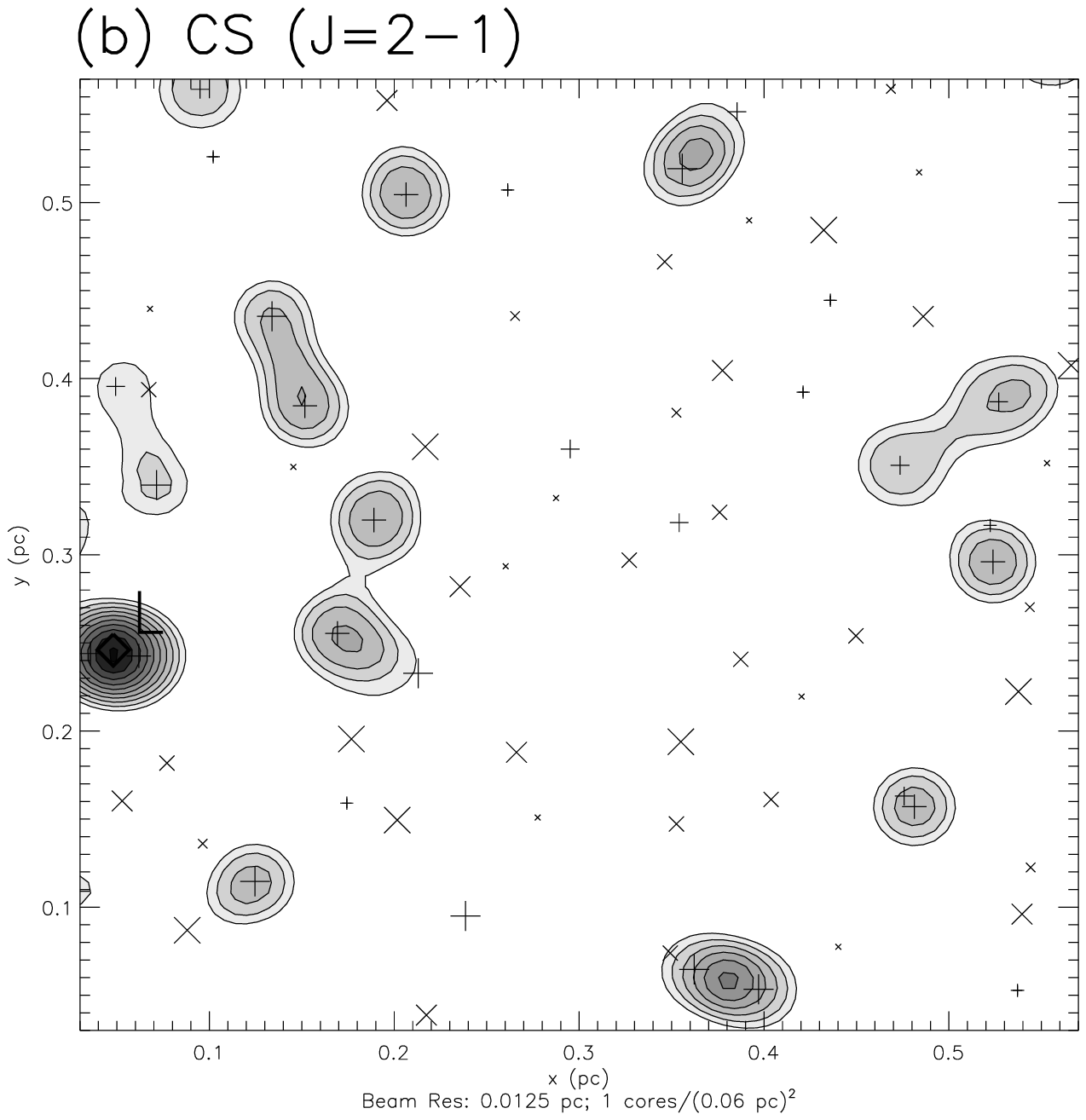}
\plotone{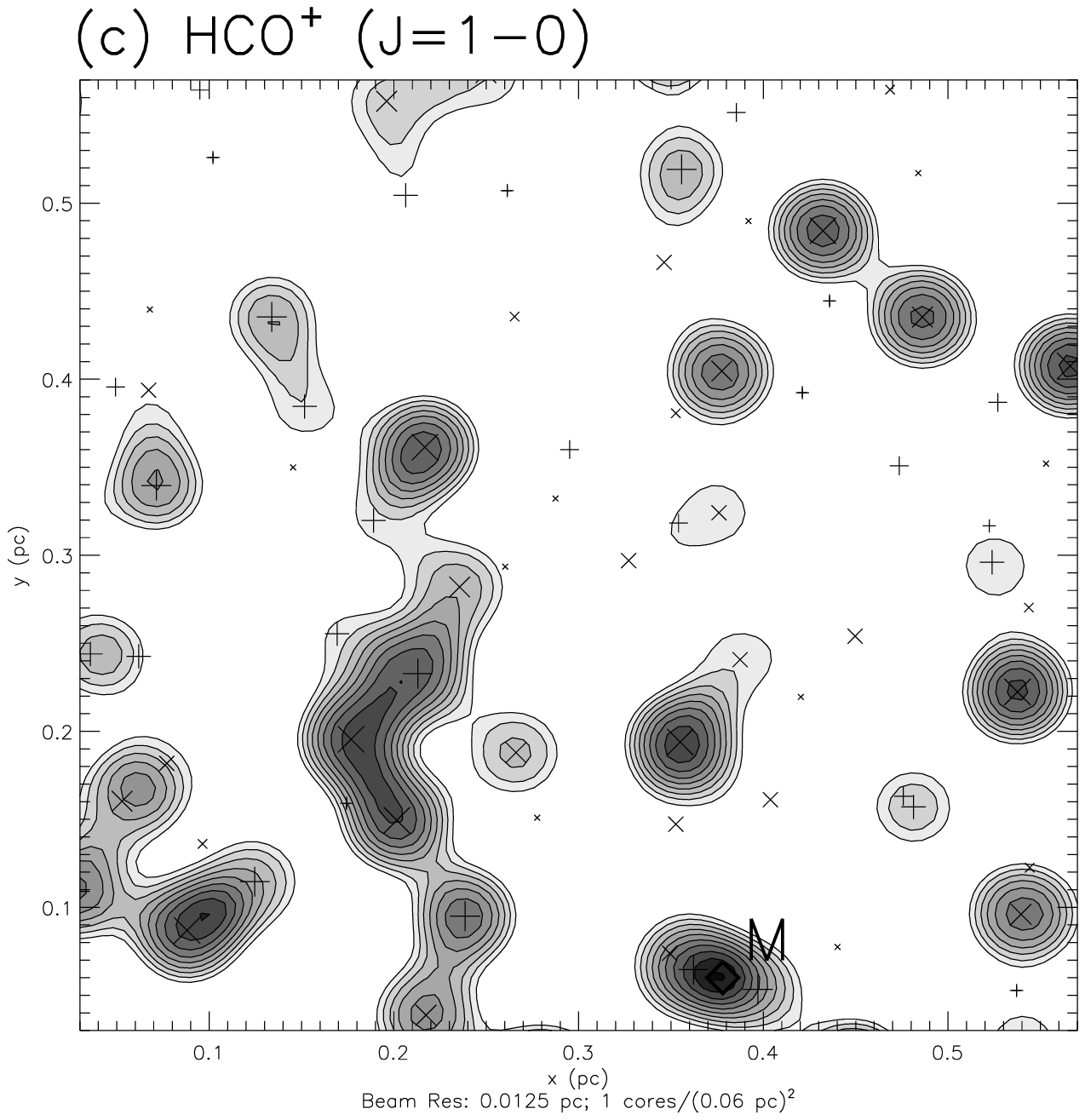}
\plotone{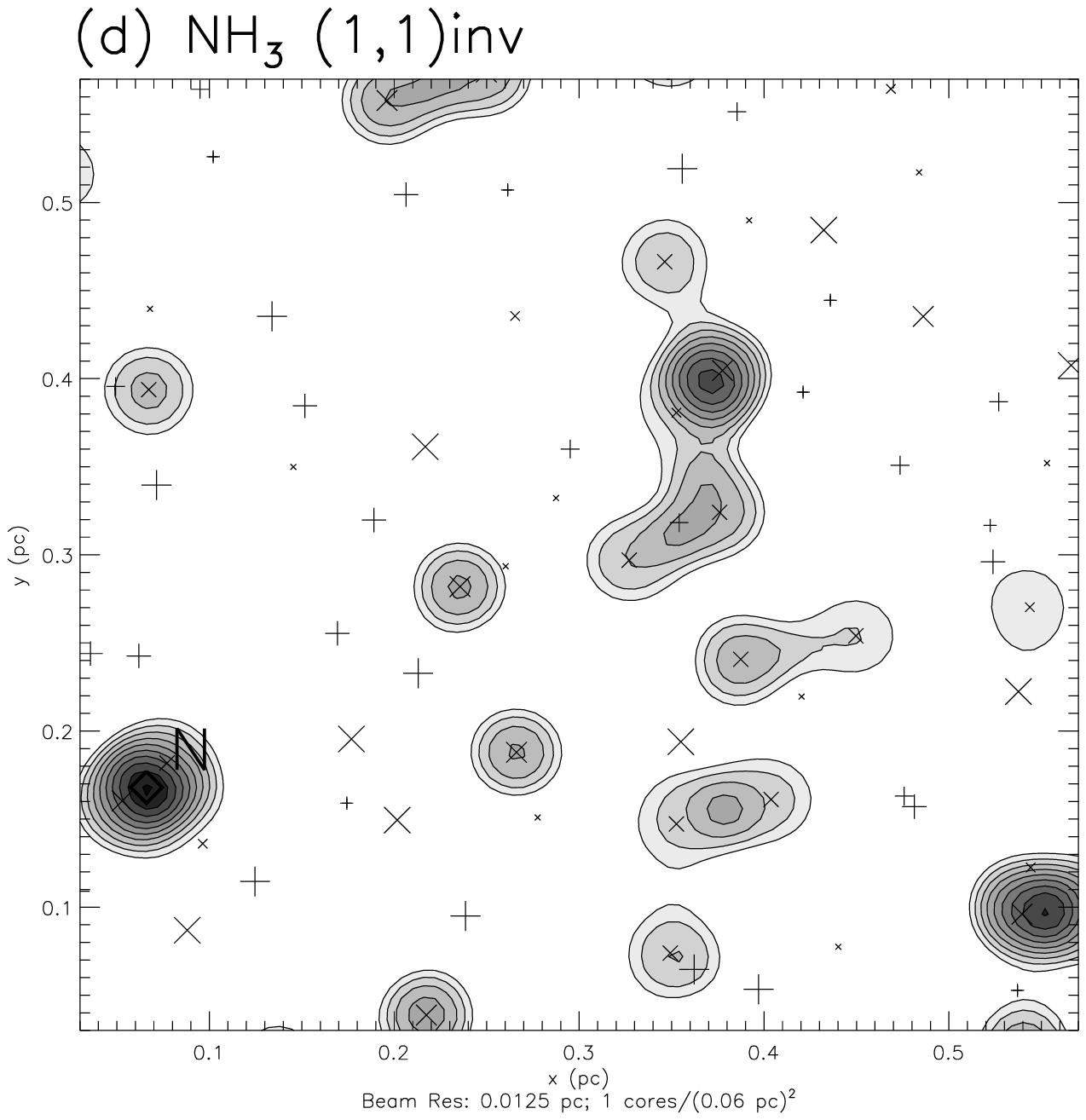}
\caption{\label{fig6} Core distribution 1, high resolution, beam 
FWHM$=0.025$ pc. Diamond symbols correspond to high resolution peaks in: (K) CO; 
(L) CS (J=2$\rightarrow$1); (M) HCO$^+$ (J=1$\rightarrow$0); 
(N) NH$_3$ (1,1)inv.}
\end{figure*}

We start by generating a quasi-random map with a very uniform distribution, and 
then stochastically move the positions of the cores by small amounts to vary 
the two statistical parameters. This results in maps whose only difference is 
in the location of cores, rather than their evolutionary status.
Three cores distributions have been considered; (1): ICD=0.524 pc, SEM=4.21\%,
(2) ICD=0.520 pc, SEM=4.85\%, and (3) ICD=0.529 pc, SEM=5.19\%.
The convolved molecular line maps for distribution (1) are shown in Figure 
\ref{fig6}.

Even with the much greater level of uniformity in distribution 1, we do not
resolve many cores individually.
We may discern, however, that for CO those peaks whose contours are very tight
tend to be the result of core associations, whereas peaks whose contours are 
broad and fairly regular tend to be single cores, close to their peak 
densities. These peaks do not tend to be as strong as those for which core 
associations are responsible. For the other molecules, most peaks tend to be 
the result of the resolution of single, fairly isolated cores.

In general we find that the apparent degree of confusion in the maps is smaller 
and the peak levels and contrasts are larger for the more random (less uniform) 
distributions. The spacing between emitting cores is the dominant factor in 
determining the morphologies of the maps. 

Using the different molecular tracers as test 
cases for these variables, we see that as in the case of the CS 
(J=2$\rightarrow$1) maps, widely spaced emitting cores produce quite tight 
peaks, and more strong peaks tend to be present with decreasing uniformity of 
general core distribution. 

Table \ref{tab6-5} shows column densities at the peak positions (for the 
strongest peak in the map) for each 
molecule. It might seem that there is a slight trend for levels to be higher 
for lower uniformity of core distributions, but this is only consistently true 
for CO. However, this does suggest firstly that the levels obtained from 
observations are quite robust with respect to the pattern of core distribution, 
and secondly that the discrepancy between the observed and modelled CS/NH$_3$ 
column densities is not likely to be resolved 
by consideration of such effects.

The observed CS J=2$\rightarrow$1 morphologies \citep{morata03a}  
most closely resemble the modelled morphologies of CO, particularly 
for the most uniform distribution, suggesting that CS cores are not highly 
clustered in L673, and that the contributing cores in that region are probably 
more numerous than is assumed in this model. The most probable explanation for 
this is that there are a number of inherently more weakly emitting cores in 
L673, which have different physical conditions from those that we have 
modelled. From the similarity between the modelled CO and the observed CS,
we may infer that the E and W peaks in the CS (J=2$\rightarrow$1) emission of 
\citet{morata03a} are more likely to be the result of core associations, whilst 
the emission at peak N may come from two underlying cores.

\section{Molecular Line Maps with Different Grain Mantle Re-injection 
Characteristics}

\begin{figure}
\epsscale{0.45}
\plotone{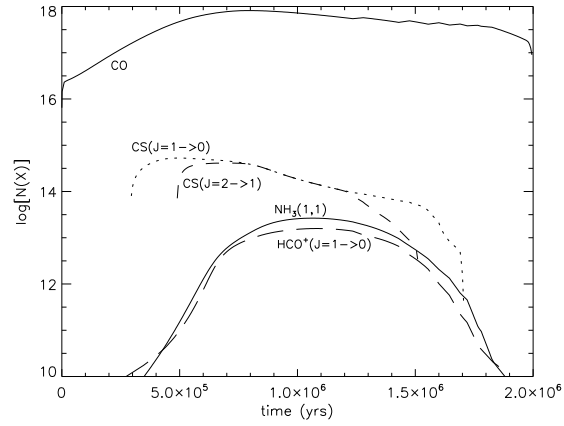}
\caption{\label{fig7} Column densities of selected transitions as 
functions of time, with $n_{\mbox{\scriptsize{{\em eff}}}}$ considerations. 
Profiles are smoothed assuming minimal re-injection.}
\end{figure}

Finally, we consider the effects of using column density profiles with a 
minimal contribution from the re-injection of dust grain mantles. This follows 
from the situation where the rate of deposition of grain mantle-bound species 
is so slow compared to the rate of destruction of injected species by the gas 
phase chemistry that there is no discernible difference in the resultant column 
density measured through the core. 
Figure \ref{fig7} shows the resultant column density profiles and can be 
compared to the `maximal' re-injection case (Fig. \ref{fig2}b).
We apply these profiles to the same basic ``sky'' map. 

\begin{figure}
\epsscale{0.35}
\plotone{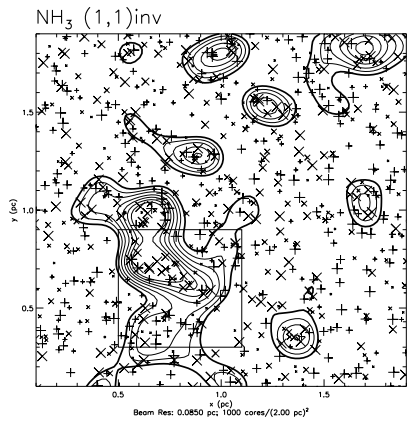}
\caption{\label{fig8} Low resolution convolved map, with minimal re-injection, 
beam FWHM$=0.17$ pc.}
\end{figure}

Of the various tracers, the most strongly affected is the NH$_3$ morphology, 
shown in figure \ref{fig8}. 
It not only assumes a smaller spatial extent but maps different material from
the maximal re-injection case - tracing regions that are very similar to those
traced by HCO$^+$ (J=1$\rightarrow$0). The relative morphologies of CS 
(J=1$\rightarrow$0), CO and NH$_3$ (1,1) are perhaps closer to those observed 
by \citet{myers91a} than in the maximal re-injection case.

However, the absolute NH$_3$ peak levels are less consistent with measured 
values. Since the absolute NH$_3$ column densities are a more robust measure, 
it is more likely that the morphologies exhibited in \citet{myers91a} etc. are 
the product of a range of chemistries engendered by a spectrum of cores in dark 
clouds, as already discussed, rather than being explained purely by the 
adoption of the minimal re-injection profile.

High resolution maps show that the differences are slight compared to the
maximal re-injection case for all species except NH$_3$, which maps very
similar regions to HCO$^+$ (J=1$\rightarrow$0) as at low resolution. 
Therefore, a test of this extreme of re-injection would be whether or not
HCO$^+$ and NH$_3$ appear to map the same regions.

\section{Summary and Conclusions}

We have constructed a molecular line contour map of a model dark cloud by
placing a large number of evolving cores randomly in space and attributing
to each a randomly determined stage of evolution. The column density data
from the chemical model of such an evolving core have been applied to each
core in the map, and all such data convolved to simulate molecular line
maps of dark cloud regions. Observational evidence of L673 from Morata et
al. (1997 and 2003) has been used to guide the choice of map parameters.

The computed low resolution morphologies of NH$_3$ and CS are found to agree 
well with the observational evidence of the spatial extent of these molecules
and with the relative positions of peaks and contours in L673. CO assumes 
a slightly smaller extent than is evidenced by the survey of \citet{myers91a}
 of 16 dark cloud cores.

The computed high resolution maps confirm that the model produces 
morphologies and peak positions similar to those in L673. The 
comparison of the synthetic maps with observational data suggests that the 
cores in L673 are randomly situated and have no particular bias in space. 

We show in Sections 4 and 5 that the general behaviour of the model is 
fairly stable against variations in the spatial distribution of cores  and 
the method of injection of ices into the gas phase during the expansion of 
the core. 

The absence of proper radiative transfer in the code was a necessary 
compromise in our calculations, but the use of effective critical emission 
densities allowed the discrimination of different molecular line 
transitions in the simulations: The convolution of data obtained from 
chemical modelling to produce maps similar to observations is striking, 
and probably unique in the history of chemical modelling of molecular 
clouds. The level of agreement between observations and models of absolute 
values of molecular column densities is good and indicates that our 
proposal to model dark clouds as an ensemble of identical transient cores 
is not merely plausible but may actually represent the true situation. 

In summary, our main conclusions from this study are as follows:
\begin{enumerate}

\item A molecular cloud model considered as a large ensemble of randomly
placed evolving cores, each with randomly allocated states of evolution,
generates molecular line maps at low and high resolution that are
remarkably similar in character to those observed in L673; the observed
structures cannot be generated by simple models of unitary homogeneous or
collapsing clouds.

\item The molecular line maps broadly reproduce morphologies, sizes of 
emitting regions, and separation in peak positions in various species, as 
found in L673.

\item In such a model, the chemistry in each core responds to the changing 
physical state of the gas as it contracts from low density to a transient 
high density state and then back to low density. For the model to generate 
molecular line maps similar to those observed, the time-scale for the 
dynamical cycle is required to be on the order of a million years. This 
conclusion is consistent with the theoretical MHD study of Falle and 
Hartquist (2002).  

\item The gas-dust interactions significantly affect the chemistry of each
core and of the cloud. For the model to match observations, the freeze-out
of gas-phase species that occurs in the denser part of the cycle must
remove a significant amount of matter (other than hydrogen and helium) from
the gas (on the order of one half) in the form of ice mantles. The ice
mantles must be returned to the gas as the core expands to a low density
state.

\item The excellent morphological match of models with observational maps of 
L673 supports the view that the cores in that cloud are randomly spaced 
and randomly evolving, rather than being structured either spatially or 
temporally. 

\item The contrast between CS and NH$_3$ distributions in L673 (and many
other dark clouds; Morata et al. 1997) is well accounted for in the model. 
This justifies the assertion of \citet{taylor96a} concerning unresolved
cores. However, the extent of the model NH$_3$ emitting region is to some
extent an artefact of the method of computing the injection of ices into
the gas.

\item Our models clearly indicate that HCO$^+$ and N$_2$H$^+$ emissions should
be co-located in the maps, contrary to the findings of \citet{morata03a}.  
However, their conclusion is that this may be an observational artefact: 
\citet{morata05a} have shown that the apparent separation of HCO$^+$ and 
N$_2$H$^+$ emissions may be caused by high optical depths in the HCO$^+$ lines.

\item Observations of co-located emission peaks in CS and HCO$^+$ are 
predicted by the model to arise from superpositions of cores at different 
stages of evolution.
\end{enumerate}

Finally, we emphasise that in the present models all the cores are assumed 
to have equal mass. It is clear, however, from the observations of 
\citet{morata03a} that a range of core masses is present in L673. As larger 
cores evolve, they may become unstable to gravitational collapse. In 
future work, we shall consider the chemical implications of core size and 
the possible relation to low mass star formation.  

\acknowledgments

R. T. G. thanks PPARC for a studentship, and the National Science Foundation 
for partial support. D. A. W. acknowledges the support of a Leverhulme Trust 
Emeritus Fellowship.

\appendix

\section{Appendix}

In \citet{garrod05a} the time- and depth-dependent density of each of the 12 
chemical reference points is defined according to
\begin{eqnarray}
\rho(\alpha,t) = \rho(0,t_{m}) \exp \left[ - \alpha ^{2} \right] \exp \left[ - 
\left( \frac{t - t_{m}}{\tau} \right)^{2} \right] 
\end{eqnarray}
where $t$ is time, $t_{m}$ is the ``peak time'' at which density is maximal 
(half way through evolution), $\alpha$ is the parametrized distance from the 
center of the core, and $\rho(0,t_{m})$ is the peak time central density. 
$\tau$ is defined according to the choice of the initial density at the core 
center (10$^3$ cm$^{-3}$) and the maximum density at the core centre, 
$\rho(0,t_{m})$ ($5 \times 10^4$ cm$^{-3}$). The distance of a parcel of gas 
from the center of the mass-conserved core is defined by
\begin{eqnarray}
& z(t) = \alpha . \Delta z(t) & \\
& \Delta z(t)= \left[ \frac{\rho(0,t_{m})}{\rho(0,t)} \right]^{\frac{1}{k}} 
\Delta z(t_{m}) &
\end{eqnarray}
where $\Delta z(t)$ is the scale width at that time, and $k$ represents the 
number of dimensions along which collapse is taking place.

If $\Delta z(t)$ represents the scale width of the core along the principal 
axis of collapse (along the line of sight), and assuming that the rates of 
collapse along the other two axes are equal (in the absence of other 
information), we may assign a core with a time-dependent radius in the plane 
of observation (representing scale width for a Gaussian function), 
$\Delta r(t)$. We may relate these quantities to the total mass of a core via:
\begin{eqnarray*}
M_{tot} \propto \rho(t) . V \propto \rho(t) . \Delta z(t) . [\Delta r(t)]^{2}
\end{eqnarray*}
Since the total mass is conserved, we have that:
\begin{eqnarray*}
\Delta z(t) . [\Delta r(t)]^{2} \propto [\rho(t)]^{-1}
\end{eqnarray*}
Therefore, eliminating $\Delta z(t)$ using equation (A3): 
\begin{eqnarray*}
\Delta r(t) \propto [\rho(t)]^{(\frac{1}{k} -1)/2}
\end{eqnarray*}
Or, eliminating $\rho(t)$:
\begin{eqnarray*}
\Delta r(t) \propto [\Delta z(t)]^{(k-1)/2}
\end{eqnarray*}
For the standard run which we use to construct the maps, $k=1.5$ 
\citep[see][]{garrod05a}, hence:
\begin{eqnarray*}
\Delta r(t) \propto [\Delta z(t)]^{1/4}
\end{eqnarray*}

\clearpage

\clearpage

\clearpage

\begin{deluxetable}{lcccc}
\tabletypesize{\scriptsize}
\tablecaption{\label{tab6-1} Properties of Line Transitions Utilised in 
Column Density Calculations}
\tablewidth{0pt}
\tablehead{
\colhead{Molecule} & \colhead{Transition} & \colhead{$\nu$} & \colhead{$n_{c}$(10 K)} & \colhead{$n_{\mbox{\scriptsize{{\em eff}}}}$ (10 K) \tablenotemark{a}} \\
\colhead{} & \colhead{} & \colhead{(GHz)} & \colhead{(cm$^{-3}$)} & \colhead{(cm$^{-3}$)}}
\startdata
CS & $J=1 \rightarrow 0$ & 49.0 & $4.6 \times 10^{4}$ & $7.0 \times 10^{3}$ \\
CS & $J=2 \rightarrow 1$ & 98.0 & $3.0 \times 10^{5}$ & $1.8 \times 10^{4}$ \\
HCO$^+$ & $J=1 \rightarrow 0$ & 89.2 & $1.7 \times 10^{5}$ & $2.4 \times 10^{3}$ \\
N$_2$H$^+$ \tablenotemark{b} & $J=1 \rightarrow 0$ & 93.2 & $1.7 \times 10^{5}$ & $2.4 \times 10^{3}$ \\
NH$_3$ & (1,1)inv & 23.7 & $1.8 \times 10^{3}$ & $1.2 \times 10^{3}$ \\
CO & --- & --- & --- & $1.0 \times 10^{3}$ \\
\enddata
\tablenotetext{a}{All values taken from \citet{evans99a}, except CO taken from 
\citet{ungerechts97a}}
\tablenotetext{b}{N$_2$H$^+$ adopts critical density values given for HCO$^+$}
\end{deluxetable}

\clearpage

\begin{deluxetable}{lccccc}
\tabletypesize{\scriptsize}
\tablecaption{\label{tab6-2} Peak Column Densities for Synthetic Maps}
\tablewidth{0pt}
\tablehead{
\colhead{Molecule} & \colhead{Transition} & \colhead{x} & \colhead{y} & \colhead{$N$[$i$] \tablenotemark{a}} & \colhead{Peak \tablenotemark{b}} \\
\colhead{} & \colhead{} & \colhead{(pc)} & \colhead{(pc)} & \colhead{(cm$^{-2}$)} &  \colhead{}}
\startdata
CO & --- & 0.66 & 0.40 & 6.0(17) & Low Res \\
 & & 0.73 & 0.46 & 2.2(18) & {\bfseries Hi -- A} \\
 & & 0.74 & 0.46 & 2.2(18) & Hi -- B \\
 & & 0.65 & 0.71 & 1.2(18) & Hi -- C \\
 & & 0.77 & 0.73 & 1.0(18) & Hi -- D  \\
\tableline
CS & J=1$\rightarrow$0 & 0.74 & 0.42 & 2.4(14) & Low Res \\
 & & 0.73 & 0.46 & 9.2(14) & {\bfseries Hi -- A} \\
 & & 0.74 & 0.46 & 9.1(14) & Hi -- B \\
 & & 0.65 & 0.71 & 2.4(14) & Hi -- C \\
 & & 0.77 & 0.73 & 2.3(14) & Hi -- D \\
\tableline
CS & J=2$\rightarrow$1 & 0.72 & 0.42 & 1.8(14) & Low Res \\
 & & 0.73 & 0.46 & 7.5(14) & Hi -- A \\
 & & 0.74 & 0.46 & 7.6(14) & {\bfseries Hi -- B} \\
 & & 0.65 & 0.71 & 2.4(14) & Hi -- C \\
 & & 0.77 & 0.73 & 1.1(14) & Hi -- D \\
\tableline
HCO$^+$ & J=1$\rightarrow$0 & 0.66 & 0.72 & 7.5(12) & Low Res \\
 & & 0.73 & 0.46 & 2.4(13) & Hi -- A \\
 & & 0.74 & 0.46 & 2.5(13) & Hi -- B \\
 & & 0.65 & 0.71 & 2.8(13) & {\bfseries Hi -- C} \\
 & & 0.77 & 0.73 & 1.2(13) & Hi -- D \\
\tableline
NH$_3$ & (1,1)inv & 0.78 & 0.70 & 7.9(13) & Low Res \\
 & & 0.73 & 0.46 & 1.4(14) & Hi -- A \\
 & & 0.74 & 0.46 & 1.2(14) & Hi -- B \\
 & & 0.65 & 0.71 & 4.7(13) & Hi -- C \\
 & & 0.77 & 0.73 & 3.1(14) & {\bfseries Hi -- D} \\
\tableline
N$_2$H$^+$ & J=1$\rightarrow$0 & 0.68 & 0.72 & 3.6(11) & Low Res \\
 & & 0.73 & 0.46 & 9.3(11) & Hi -- A \\
 & & 0.74 & 0.46 & 9.6(11) & Hi -- B \\
 & & 0.65 & 0.71 & 1.3(12) & {\bfseries Hi -- C} \\
 & & 0.77 & 0.73 & 7.4(11) & Hi -- D \\
\enddata
\tablenotetext{a}{ $a(b)=a \times 10^{b}$}
\tablenotetext{b}{High resolution peaks in bold correspond to the peak in the 
given transition. Low resolution peak positions are different for each 
transition.}
\end{deluxetable}

\clearpage

\begin{deluxetable}{lccc}
\tabletypesize{\scriptsize}
\tablecaption{\label{tab6-3} Peak Column Densities from Low and High Resolution 
Morata et al. Observations}
\tablewidth{0pt}
\tablehead{
\colhead{Molecule} & \colhead{Transition} & \colhead{$N$[$i$] \tablenotemark{a}} & \colhead{Peak \tablenotemark{b}} \\
\colhead{} & \colhead{} & \colhead{(cm$^{-2}$)} & \colhead{}}
\startdata
CS & J=1$\rightarrow$0 & 2.5(13) & M97 -- Low Res \\
\tableline
CS & J=2$\rightarrow$1 & 1.20(12) -- 2.69(12) & M03 -- S \\
 & & 2.53(12) -- 8.48(12) & {\bfseries M03 -- E} \\
 & & 3.72(12) -- 9.96(12) & {\bfseries M03 -- W} \\
 & & 2.71(12) -- 7.22(12) & M03 -- N \\
\tableline
HCO$^+$ & J=1$\rightarrow$0 & $<$4.73(11) & M03 -- S \\
 & & $<$3.60(11) & M03 -- E \\
 & & 4.7(11) -- 7.7(11) & M03 -- W \\
 & & 7.7(11) -- 1.91(12) & {\bfseries M03 -- N} \\
\tableline
NH$_3$ & (1,1)inv & $\geq$2.2(14) & M97 -- Low Res \\
\tableline
N$_2$H$^+$ & J=1$\rightarrow$0 & 2.2(11) -- 3.1(11) & {\bfseries M03 -- S} \\
 & & 1.7(11) -- 2.4(11) & M03 -- E \\
 & & $<$2.0(11) & M03 -- W \\
 & & $<$1.6(11) & M03 -- N \\
\enddata
\tablenotetext{a}{ $a(b)=a \times 10^{b}$}
\tablenotetext{b}{Values are taken from \citet{morata97a} and 
\citet{morata03a}. High resolution peaks in bold correspond to the peak in 
the given transition. Low resolution peak positions are different for each 
transition.}
\end{deluxetable}

\clearpage

\begin{deluxetable}{lccccc}
\tabletypesize{\scriptsize}
\tablecaption{\label{tab6-5} Peak Column Densities for Maps of Various Levels 
of Uniformity of Core Distribution}
\tablewidth{0pt}
\tablehead{
\colhead{Molecule} & \colhead{Transition} & \colhead{Distribution} & \colhead{x} & \colhead{y} & \colhead{$N$[$i$] \tablenotemark{a}} \\
\colhead{} & \colhead{} & \colhead{} & \colhead{(pc)} & \colhead{(pc)} & \colhead{(cm$^{-2}$)}}
\startdata
CO & --- & 1 & 0.048 & 0.240 & 1.4(18) \\
(Peak K) & & 2 & 0.186 & 0.456 & 1.5(18) \\
 & & 3 & 0.390 & 0.210 & 2.0(18) \\
\tableline
CS & J=2$\rightarrow$1 & 1 & 0.048 & 0.246 & 6.8(14) \\
(Peak L) & & 2 & 0.090 & 0.030 & 6.1(14) \\
 & & 3 & 0.390 & 0.210 & 1.0(15) \\
\tableline
HCO$^+$ & J=1$\rightarrow$0 & 1 & 0.378 & 0.060 & 1.7(13) \\
(Peak M) & & 2 & 0.384 & 0.480 & 2.5(13) \\
 & & 3 & 0.066 & 0.486 & 2.0(13) \\
\tableline
NH$_3$ & (1,1)inv & 1 & 0.066 & 0.168 & 2.9(14) \\
(Peak N) & & 2 & 0.384 & 0.474 & 3.2(14) \\
 & & 3 & 0.480 & 0.294 & 3.0(14) \\
\enddata
\tablenotetext{a}{ $a(b)=a \times 10^{b}$}
\end{deluxetable}

\end{document}